\documentclass[aps,prd,showpacs,floatfix,eqsecnum]{revtex4}
\usepackage{amsmath,latexsym,graphicx,amsbsy,natbib}

\DeclareMathOperator{\Tr}{Tr}

\usepackage[usenames]{color}

\newcommand{\fbend}{\ensuremath{f_b}}
\newcommand{\fstar}{\ensuremath{f_*}}
\newcommand{\fyr}{f_\mathrm{1yr}}
\newcommand{\SNR}{\ensuremath{\left\langle \rho \right\rangle}}
\newcommand{\HD}{\left(\sum_{IJ} \chi_{IJ}^2\right)^{1/2}}
\newcommand{\Pgw}{\ensuremath{P_\mathrm{gw}}}
\newcommand{\Agw}{\ensuremath{A_\mathrm{gw}}}
\newcommand{\F}{\ensuremath{{_2F_1}}}
\newcommand{\res}{\ensuremath{\mathbf{r}}}
\newcommand{\Sigmatrix}{\ensuremath{{\pmb\Sigma}}}
\newcommand{\Smatrix}{\ensuremath{{\mathbf{S}}}}
\newcommand{\Pmatrix}{\ensuremath{{\mathbf{P}}}}
\newcommand{\R}{\ensuremath{\mathbf{R}}}
\newcommand{\Msun}{\ensuremath{M_\odot}}
\newcommand{\Msol}{\ensuremath{M_\odot}}

\begin{document}

\title{Supermassive Black Hole Binary Environments: Effects on the
  Scaling Laws and Time to Detection for the Stochastic Background}

\author{S.~J.~Vigeland}
\author{X.~Siemens}
\affiliation{Department of Physics, University of
  Wisconsin--Milwaukee, Milwaukee, WI 53201, USA}

\date{\today}

\begin{abstract}
One of the primary gravitational wave (GW) sources for pulsar timing arrays (PTAs) 
is the stochastic background formed by supermassive black holes binaries (SMBHBs). 
In this paper, we investigate how the environments of SMBHBs will effect the sensitivity of PTAs 
by deriving scaling laws for the signal-to-noise ratio (SNR) of the optimal cross-correlation statistic. 
The presence of gas and stars around SMBHBs will accelerate the merger at large distances, 
depleting the GW stochastic background at low frequencies. 
We show that environmental interactions may delay detection 
by a few years or more, depending on the PTA configuration and the 
frequency at which the dynamical evolution transitions from being dominated 
by environmental effects to GW-dominated.
\end{abstract}
\pacs{04.80.Nn, 95.55.Ym, 97.60.Gb, 98.62.Js}
\maketitle

\section{Introduction}

Efforts are currently underway to detect low-frequency gravitational waves (GWs) 
with pulsar timing arrays (PTAs). Current sensitivities and 
modeling of GW sources at nano-Hertz frequencies point to an imminent detection of GWs \citep{2016ApJ...819L...6T}. 
PTAs are sensitive to GWs with frequencies $\sim 10^{-7} - 10^{-9} \; \mathrm{Hz}$, 
and one of the most promising sources in this frequency range is a stochastic background 
from supermassive black hole binaries (SMBHBs), which form in galaxy mergers 
\citep{1995ApJ...446..543R, 2003ApJ...583..616J, 2003ApJ...590..691W}. Assuming 
these binaries evolve solely due to GW emission, the stochastic background can be 
described by a power-law \citep{2016ApJ...821...13A}:
\begin{equation}
	h_c(f) = A_\mathrm{gw} \left( \frac{f}{\fyr} \right)^\alpha \,,
	\label{eq:GWbackground_powerlaw}
\end{equation}
where $\fyr \equiv (1 \: \mathrm{yr})^{-1}$, $A_\mathrm{gw}$ is the dimensionless 
GW amplitude at $f = \fyr$, and $\alpha=-2/3$.

However, observations of supermassive black holes (SMBHs) suggest they 
are often surrounded by gas and stars, 
which will alter the GW spectrum by accelerating the orbital decay. 
Additionally, SMBHBs will not merge in less than a Hubble time if they only evolve due to 
GW emission; the energy loss must be accelerated through other mechanisms 
(the so-called ``last-parsec problem'') 
\citep{2003AIPC..686..201M, 2012arXiv1211.5377M, 2014MNRAS.442...56R, 2014ApJ...785..163V, 2015PhRvD..91h4055S}.
\citet{2015PhRvD..91h4055S} introduced a new parametrization of the GW strain 
to account for binaries evolving due to physical processes besides just GW emission. 
In general, the GW strain from many circular SMBHBs can be written as \citep{2001astro.ph..8028P}
\begin{equation}
	h_c^2(f) = \int_0^\infty dz \int_0^\infty d\mathcal{M} \frac{d^3N}{dz \: d\mathcal{M} \: dt} \frac{dt}{d \ln f} h^2(f) \,,
	\label{eq:GWbackground}
\end{equation}
where $z$ is the redshift, $N$ is the number of inspiraling binaries, 
$t$ is time measured in the binary's rest frame, and $\mathcal{M}$ is a combination of the 
black hole masses $M_1$ and $M_2$ known as the chirp mass,
\begin{equation}
	\mathcal{M} \equiv \frac{(M_1 M_2)^{3/5}}{(M_1+M_2)^{1/5}} \,.
\end{equation}
The GW strain of a single circular binary $h(f)$ is \citep{1963PhRv..131..435P, 2003ApJ...583..616J}
\begin{equation}
	h(f) = \sqrt{\frac{32}{5}} \frac{\mathcal{M}^{5/3}}{D_L} (\pi f)^{2/3} \,,
\end{equation}
where $D_L$ is the luminosity distance to the source. 
If the binaries are evolving solely due to GW emission, then the frequency evolution is given by
\begin{equation}
	\frac{dt}{d \ln f} = \frac{5}{96 \pi^{8/3}} \mathcal{M}^{-5/3} f^{-8/3} \,.
\end{equation}
Substituting this into Eq.~\eqref{eq:GWbackground} gives 
\begin{equation}
	h_c^2(f) = \frac{1}{3 \pi^{4/3}} f^{-4/3} \int_0^\infty dz \int_0^\infty d\mathcal{M} \frac{d^3N}{dz \: d\mathcal{M} \: dt} \frac{\mathcal{M}^{5/3}}{D_L^2} \,.
\end{equation}
Putting this into the same form as Eq.~\eqref{eq:GWbackground_powerlaw}, we see that 
the GW spectrum amplitude $\Agw$ is
\begin{equation}
	\Agw = \fyr^\alpha \left[ \frac{1}{3 \pi^{4/3}} \int_0^\infty dz \int_0^\infty d\mathcal{M} \frac{d^3N}{dz \: d\mathcal{M} \: dt} \frac{\mathcal{M}^{5/3}}{D_L^2} \right]^{1/2} \,.
\end{equation}

As introduced by \citet{2015PhRvD..91h4055S}, 
we can account for astrophysical mechanisms that evolve the binaries 
by parametrizing the frequency evolution as
\begin{equation}
	\frac{dt}{d \ln f} = \frac{1}{f} \left[ \sum_i \left(\frac{df}{dt}\right)_i \right]^{-1} \,,
\end{equation}
where each mechanism is associated with a different frequency evolution $(df/dt)_i$.  
For binaries that evolve due to GW emission and one additional mechanism, the 
GW strain can be written as a broken power-law \cite{2015PhRvD..91h4055S}:
\begin{equation}
	h_c(f) = A_\mathrm{gw} \frac{(f/\fyr)^{\alpha}}{\left[1+(\fbend/f)^\kappa\right]^{1/2}} \,,
	\label{eq:GWbackground_bend}
\end{equation}
where $\kappa$ and $\fbend$ parametrize the process other than GW emission 
that drive the binary evolution. For stellar scattering, $\kappa=10/3$, and for 
alpha-disk binary interactions, $\kappa=7/3$. 
The bend frequency $\fbend$ depends on the density of stars or gas near the SMBHs. 
At high frequencies, the broken-power-law and power-law spectra are identical, but at 
low frequencies there can be significantly less power. 
\citet{2016ApJ...819L...6T} used simulations to estimate the time to detection for a 
power-law spectrum and a broken-power-law spectrum with 
and concluded that a turnover at $\fbend=3\times10^{-9} \: \mathrm{Hz}$ 
does not significantly delay detection.

In this paper, we derive scaling laws for the signal-to-noise ratio (SNR) 
of the optimal cross-correlation statistic 
in order to investigate how this loss of power at low frequencies will affect the 
sensitivity of PTAs, 
depending on the spectrum parameters and PTA configuration. 
The optimal cross-correlation statistic 
is an estimator of the GW stochastic background 
amplitude in the weak-signal regime \citep{2009PhRvD..79h4030A}. 
\citet{2013CQGra..30v4015S} and \citet{2015PhRvD..91d4048C} used the 
optimal cross-correlation statistic to derive analytic scaling laws 
for the SNR for a power-law stochastic background. 
This allowed them to determine how PTAs can be optimized 
in terms of the number of pulsars, observation time, observing cadence, and the 
pulsar timing RMS in order to maximize their sensitivity to 
the stochastic background, and to estimate the time to detection for various 
detector configurations. 
They concluded that if the detector is operating in the weak-signal regime, 
where the GW signal is below the white noise at all frequencies, 
the best way to improve the SNR is to decrease the pulsars' timing RMS. 
As PTAs become sensitive to lower frequencies, they transition into the 
so-called ``intermediate-signal regime,'' where the GW signal is below the white 
noise at high frequencies but above the white noise at low frequencies. 
In this regime, decreasing the timing RMS does not significantly 
improve the SNR, and the best strategy is to add new pulsars to the PTA. 
Our goal is to derive similar relationships in order to determine how 
the spectral shape affects time to detection and the optimal PTA configuration.

This paper is organized as follows. 
In Sec.~\ref{sec:optstat} we review the optimal cross-correlation statistic and 
how to compute the SNR from it. 
In Sec.~\ref{sec:regimes} we discuss how the properties of the PTA 
and parameters of the GW spectrum affect whether our PTA is operating 
in the weak-signal or intermediate-signal regime. We derive scaling laws 
for the SNR in the weak-signal and intermediate signal regime in 
Sec.~\ref{sec:scalinglaws}. In Sec.~\ref{sec:numerical} we compare the scaling laws 
derived in the previous section to numerical calculations of the SNR for various 
combinations of the GW spectrum parameters and PTA parameters. 
We summarize our results and discuss future work in Sec.~\ref{sec:conclusions}.

\section{Optimal cross-correlation statistic}
\label{sec:optstat}

The optimal cross-correlation statistic introduced by \citet{2009PhRvD..79h4030A} 
is an estimator for the amplitude of an isotropic GW stochastic background in the weak-signal regime. 
To derive the optimal cross-correlation statistic, we start from the likelihood function for the timing residuals: 
\begin{equation}
	p(\res|\phi) = \frac{\exp\left[-\frac{1}{2} \res^T \; \Sigmatrix^{-1}(\phi) \; \res \right]}{\sqrt{\det[2\pi \Sigmatrix(\phi)]}} \,,
\end{equation}
where $\phi$ is a set of parameters that describe the noise, and 
\res\ is an array of all the timing residuals, 
\begin{equation}
	\res = \left[ \begin{array}{c} \res_1 \\ \res_2 \\ \vdots \\ \res_M \end{array} \right] \,,
\end{equation}
where $M$ is the number of pulsars in our PTA. 
The covariance matrix of the residuals \Sigmatrix\ is given by
\begin{equation}
	\Sigmatrix = \R \, \Sigmatrix_n \, \R^T \,,
\end{equation}
where $\R$ is the matrix that accounts for the fitting to the timing model, 
and $\Sigmatrix_n$ is the pre-fit covariance matrix. \Sigmatrix\ can be written as
\begin{equation}
	\Sigmatrix = \left[ \begin{array}{cccc} \Pmatrix_1 & \Smatrix_{12} & \hdots & \Smatrix_{1M}  \\
							\Smatrix_{21} & \Pmatrix_2 & \hdots & \Smatrix_{2M} \\
							\vdots & \vdots & \ddots & \vdots \\
							\Smatrix_{M1} & \Smatrix_{M2} & \hdots & \Pmatrix_M \end{array} \right] \,,
\end{equation}
where $I$ indicates the pulsar number, 
and $\Pmatrix_I$ are the auto-covariance matrix and 
$\Smatrix_{IJ}$ are the cross-covariance matrices 
for each pulsar and pulsar pair:
\begin{eqnarray}
	\Pmatrix_I &=& \left\langle \res_I \res_I^T \right\rangle \,, \\
	\Smatrix_{IJ} &=& \left. \left\langle \res_I \res_J^T \right\rangle \right|_{I \neq J} \,.
\end{eqnarray}
The auto-covariance and cross-covariance matrices are related to the GW power spectrum by
\begin{eqnarray}
	\left\langle \res_I \res_I^T \right\rangle_{ij} &=& \R_I \left[ \int_{-\infty}^\infty df \; e^{2\pi i f \tau_{ij}} P_I(f) \right] \R_I^T \,, \\
	\left. \left\langle \res_I \res_J^T \right\rangle \right|_{ij} &=& \chi_{IJ} \R_I \left[ \int_{-\infty}^\infty df \; e^{2\pi i f \tau_{ij}} P_\mathrm{gw}(f) \right] \R_J^T \,,
\end{eqnarray}
where $i,j$ index the matrix elements, $\tau_{ij} \equiv |t_i - t_j|$ is the time interval between 
two observations, 
$\chi_{IJ}$ are the Hellings-Downs coefficients, 
$P_\mathrm{gw}(f)$ is the GW power spectrum, 
and $P_I(f)$ is the power spectrum for the $I$th pulsar. 
Assuming each pulsar's timing residuals only contain white noise and a GW, 
$P_I(f)$ can be written as
\begin{equation}
	P_I(f) = P_\mathrm{gw}(f) + 2 \sigma^2 \Delta t \,,
\end{equation}
where $\sigma$ is the residual RMS and $\Delta t$ is the time between observations.

As shown in \citet{2009PhRvD..79h4030A}, in the weak-signal regime analytically maximizing the 
likelihood ratio over amplitudes yields the optimal statistic,
\begin{equation}
	\hat{A}^2 = \frac{\sum_{IJ} \res_I^T \Pmatrix_I^{-1} \tilde{\Smatrix}_{IJ} \Pmatrix_J^{-1} \res_J}{\sum_{IJ} \Tr \left( \Pmatrix_I^{-1} \tilde{\Smatrix}_{IJ} \Pmatrix_J^{-1} \tilde{\Smatrix}_{JI} \right) } \,,
\end{equation}
where $\tilde{\Smatrix}_{IJ}$ is the amplitude-independent cross-correlation matrix,
\begin{equation}
	\Agw^2 \tilde{\Smatrix}_{IJ} = \Smatrix_{IJ} \,,
\end{equation}
and the normalization factor has been chosen so that 
$\langle \hat{A}^2 \rangle = \Agw^2$.
If there is no GW signal or the signal is weak, the expectation value of $\hat{A}$ vanishes and the 
standard deviation is
\begin{equation}
	\sigma_0 = \left[ \sum_{IJ} \Tr \left( \Pmatrix_I^{-1} \tilde{\Smatrix}_{IJ} \Pmatrix_J^{-1} \tilde{\Smatrix}_{JI} \right) \right]^{-1/2} \,.
\end{equation}
For a particular measurement of the optimal statistic, the signal-to-noise (S/N) ratio is
\begin{equation}
	\hat{\rho} = \frac{\hat{A}^2}{\sigma_0} = \frac{\sum_{IJ} \res_I^T \Pmatrix_I^{-1} \tilde{\Smatrix}_{IJ} \Pmatrix_J^{-1} \res_J}{\left[ \sum_{IJ} \Tr \left( \Pmatrix_I^{-1} \tilde{\Smatrix}_{IJ} \Pmatrix_J^{-1} \tilde{\Smatrix}_{JI} \right) \right]^{1/2}} \,.
\end{equation}
The expectation value of the SNR is
\begin{equation}
	\SNR = \Agw^2 \left[ \sum_{IJ} \Tr\left( \Pmatrix_I^{-1} \tilde{\Smatrix}_{IJ} \Pmatrix_J^{-1} \tilde{\Smatrix}_{JI} \right) \right]^{1/2} \,.
\end{equation}
In the frequency domain, we can express the SNR as 
\citep{2009PhRvD..79h4030A, 2013CQGra..30v4015S}
\begin{equation}
	\SNR = \left[ 2T \sum_{IJ} \chi_{IJ}^2 \int_{f_L}^{f_H} df \frac{P_\mathrm{gw}(f)^2}{P_I(f) P_J(f)} \right]^{1/2} \,.
\end{equation}
Assuming that all pulsars have the same noise characteristics, this becomes
\begin{equation}
	\SNR = \HD \left\{ 2T \int_{f_L}^{f_H} df \frac{P_\mathrm{gw}(f)^2}{\left[P_\mathrm{gw}(f) + 2\sigma^2 \Delta t \right]^2} \right\}^{1/2} \,.
	\label{eq:SNR}
\end{equation}

\section{The weak-signal and intermediate-signal scaling regimes}
\label{sec:regimes}

As discussed in \citet{2013CQGra..30v4015S} and \citet{2015PhRvD..91d4048C}, 
we can get an intuitive understanding of how the sensitivity of PTAs depend upon their 
configuration and the form of the spectrum by integrating Eq.~\eqref{eq:SNR}. 
We consider two regimes: the weak-signal regime, for which the intrinsic white noise dominates at all frequencies, and the intermediate-signal regime, for which the white noise dominates at high frequencies, 
but the GW signal dominates at low frequencies.

We write the GW power spectral density as
\begin{equation}
	P(f) = \frac{bf^{-\gamma}}{1+(\fbend/f)^\kappa} \,,
	\label{eq:Pgw}
\end{equation}
where $\gamma = 3-2\alpha$, and $b = A_\mathrm{gw}^2/(12\pi^2 \fyr^{2\alpha})$. This can be approximated by the piecewise function:
\begin{eqnarray}
	P(f) = \left\{ \begin{array}{lc} bf^{-\gamma} & f > \fbend \,, \\
						b \fbend^{-\kappa} f^{-(\gamma-\kappa)} & f < \fbend \,. \end{array} \right.
	\label{eq:Pgw_piecewise}
\end{eqnarray} 
The parameter $\kappa$ depends on the dominant astrophysical mechanism 
responsible for the solution to the last parsec problem, with 
$\kappa=10/3$ corresponding to stellar scattering, 
and $\kappa=7/3$ to $\alpha$-disk binary interactions.

The bend frequency \fbend\ depends on the density of matter near the SMBHBs. 
For stellar scattering, \fbend\ is related to astrophysical parameters according to \citep{2016ApJ...821...13A}
\begin{equation}
	\fbend = (3.13 \: \mathrm{nHz}) \; \left(\frac{\sigma}{200 \; \mathrm{km}/\mathrm{s}}\right)^{-3/10} \left( \frac{\rho}{10^3 \; \Msun \, \mathrm{pc}^{-3}} \right)^{3/10} \left(\frac{H}{15}\right)^{3/10} \left( \frac{M}{10^8 \; \Msun} \right)^{-2/5} q_r^{-3/10} \,,
\end{equation}
where $\sigma$ is the stellar velocity dispersion, $\rho$ is the stellar mass density, 
$H$ is a dimensionless hardening parameter, $M$ is the total mass of the binary, 
$q=M_2/M_1$ is the mass ratio, and $q_r = q/(1+q)^2$. 
Expected values for $\sigma$, $\rho$, $H$, and $M$ predict $\fbend \sim 3 \: \mathrm{nHz}$, 
which coincides with the smallest value of \fbend\ that we consider here. 
We also consider a higher value of \fbend\ because, as we show below, 
values of \fbend\ below a few nHz will have almost no effect on PTAs. 
This is due to the fact that long observation times are required 
for PTAs to be sensitive to those low frequencies. 
Higher values of \fbend\ require higher stellar densities -- 
assuming black hole masses of $\sim 10^8 - 10^9 \: \Msol$, 
a bend frequency of $\fbend = 10^{-8} \; \mathrm{Hz} = 1/(3 \; \mathrm{yr})$ 
requires $\rho \sim 10^4 - 3 \times10^5 \: \Msun \, \mathrm{pc}^{-3}$. 
Based on observations of massive elliptical galaxies, 
we expect the stellar density to be $\sim10 - 10^4 \: \Msol \mathrm{pc}^{-3}$ 
\citep{2007MNRAS.379..956D}, 
which corresponds to $\fbend \sim 5\times10^{-10} - 5 \times 10^{-9} \: \mathrm{Hz}$ 
for black hole masses of $\sim 10^8 - 10^9 \: \Msol$. 
For gas-driven dynamics, 
\begin{equation}
	\fbend \approx (0.144 \: \mathrm{nHz}) \; M^{-17/14} q_r^{-6/7} \dot{M}_1^{3/7} a_0^{3/10} \,,
\end{equation}
where $\dot{M}_1$ is the accretion rate into the primary black hole and 
$a_0$ is the orbital separation at which the mass of the enclosed disk is equal to the mass of 
the secondary black hole,
\begin{equation}
	a_0 = (3\times10^3) \left(\frac{\alpha}{10^{-2}}\right)^{4/7} \left( \frac{M_2}{10^6 \; \Msun} \right)^{5/7} \left( \frac{M_1}{10^8 \; \Msun} \right)^{-11/7} \left( 100 \frac{\dot{M}_1}{\dot{M}_E} \right)^{-3/7} r_S \,,
\end{equation}
where $\alpha$ is the disk viscosity parameter; 
$M_1$ and $M_2$ are the masses of the primary and secondary black holes, respectively; 
$\dot{M}_E$ is the Eddington accretion rate of the primary black hole; and 
$r_S = 2GM_1/c^2$ is the Schwarzschild radius of the primary black hole \citep{2016ApJ...821...13A}. 
The bend frequency is most sensitive to the accretion rate $\dot{M}_1$, with higher values of 
$\fbend$ implying larger $\dot{M}_1$. Currently, estimates of $\dot{M}_1$ vary over several orders of magnitude from $10^{-3} \; \Msun \: \mathrm{yr}^{-1}$ up to $1 \; \Msun \: \mathrm{yr}^{-1}$ 
\citep{2001ApJ...547..731D, 2002ApJ...567L...9A, 2015MNRAS.448.3603D, 2016MNRAS.455.1989G}, 
which corresponds to $\fbend \sim 6\times10^{-10} - 5\times10^{-8} \: \mathrm{Hz}$. 
The two values for \fbend\ that we consider in this paper, 
$\fbend = 3\times10^{-9} \: \mathrm{Hz}$ and $\fbend = 10^{-8} \: \mathrm{Hz}$, 
lie within this expected range. Furthermore, our chosen values for \fbend\ 
allow us to illustrate a wide range of possible behaviors of the SNR.

\begin{figure}
	\includegraphics[width=0.9\columnwidth]{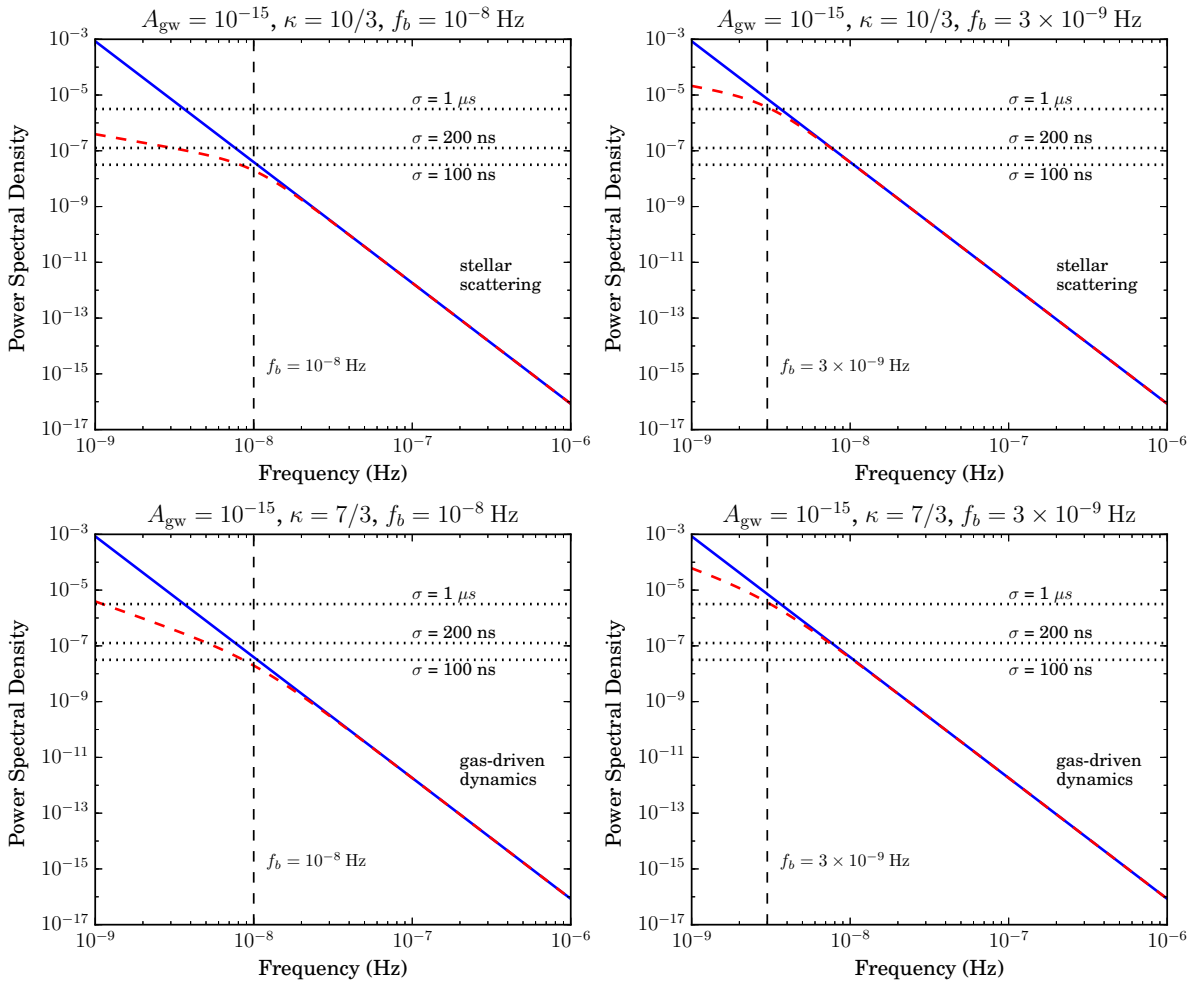}
	\caption{The power spectral density for a power-law GW spectrum 
			with $\gamma=13/3$ (solid blue line) 
			and a broken-power-law spectrum (dashed red line). 
			The top row shows spectra with $\kappa=10/3$ (stellar scattering), 
			while the bottom row shows spectra with $\kappa=7/3$ (gas-driven dynamics). 
			We also plot two different values of \fbend: 
			$\fbend = 10^{-8} \: \mathrm{Hz} = 1/(3 \: \mathrm{yr})$ (left column) 
			and $\fbend = 3 \times 10^{-9} \: \mathrm{Hz} = 1/(10 \: \mathrm{yr})$ (right column). 
			The dotted horizontal lines indicated the power spectral density of the white noise 
			for $\sigma = 1\;\mathrm{\mu s}$ (top line), $\sigma = 200\;\mathrm{ns}$ (middle line), 
			and $\sigma = 100\;\mathrm{ns}$ (bottom line). The dashed vertical line indicates 
			the bend frequency $\fbend$.}
	\label{fig:GWspectrum}
\end{figure}

Figure~\ref{fig:GWspectrum} shows the GW spectrum for several combinations of $\kappa$ and \fbend, 
along with a power-law GW spectrum. 
We also plot the power spectral density of the white noise for three values of the timing RMS: 
$\sigma = 1\;\mathrm{\mu s}$, $\sigma = 200\;\mathrm{ns}$, and $\sigma = 100\;\mathrm{ns}$. 
This should be compared with, for example, 
the NANOGrav 9-yr data release, which consisted of 37 pulsars, all of which had 
residuals $<1 \; \mu\mathrm{s}$, and 12 of which had residuals $<200 \; \mathrm{ns}$ 
\citep{2015ApJ...813...65T}.

The frequency at which the GW spectum and white noise are equal is \fstar,
\begin{equation}
	2\sigma^2 \Delta t = \frac{b\fstar^{\gamma}}{1+(f_b/\fstar)^\kappa} \,.
\end{equation}
Although we cannot write down an exact analytic expression for \fstar, 
in the limit that $\fstar \ll \fbend$ or $\fstar \gg \fbend$, \fstar\ is given by
\begin{eqnarray}
	\fstar = \left\{ \begin{array}{lc} \left(\frac{b}{2\sigma^2 \Delta t}\right)^{1/\gamma} & \fstar \gg \fbend \,, \\
						\left(\frac{b \fbend^{-\kappa}}{2\sigma^2 \Delta t}\right)^{1/(\gamma-\kappa)} & \fstar \ll \fbend \,. \end{array} \right.
\end{eqnarray}
We transition from the weak-signal regime to the intermediate-signal regime when the lowest frequency bin $f_L = 1/T$ is less than \fstar. We can use this requirement to get a relationship between the observation time and the timing RMS,
\begin{equation}
	\sigma < \left[ \left(\frac{1}{2 \Delta t} \right) \frac{bT^{-\gamma}}{1+(\fbend T)^\kappa}\right]^{1/2} \,,
\end{equation}
or, using the piecewise form of $\fstar$,
\begin{eqnarray}
	\sigma < \left\{ \begin{array}{lc} \left(\frac{b}{2 \Delta t}\right)^{1/2} T^{\gamma/2} & T \ll \fbend^{-1} \,, \\
						\left(\frac{b \fbend^{-\kappa}}{2 \Delta t}\right)^{1/2} T^{(\gamma-\kappa)/2} & T \gg \fbend^{-1} \,. \end{array} \right.
\end{eqnarray}

\begin{figure}
	\includegraphics[width=0.9 \columnwidth]{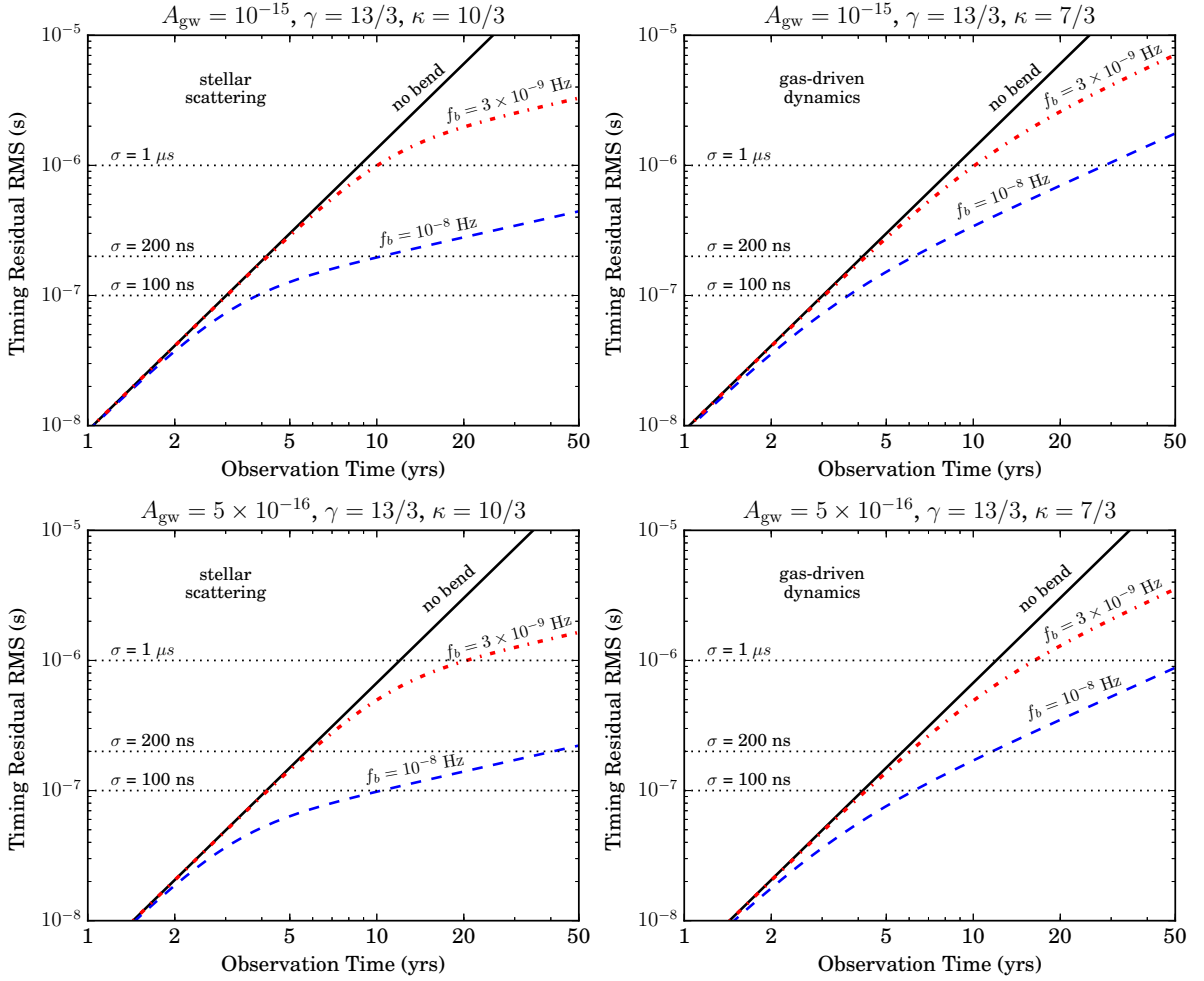}
	\caption{Timing RMS at which the GW power in the lowest frequency bin is equal to 
			the white noise level, i.e., the transition from the weak-signal regime 
			to the intermediate-signal regime, as a function of observation time. 
			We compare a power-law spectrum (solid black line), 
			a broken-power-law spectrum with 
			$\fbend=10^{-8} \: \mathrm{Hz} = 1/(3 \: \mathrm{yr})$ (dashed blue line), 
			and a broken-power-law spectrum with 
			$\fbend=3\times10^{-9}\:\mathrm{Hz} = 1/(10 \: \mathrm{yr})$ (dotted-dashed red line). 
			On the left we plot a GW spectrum with $\kappa=10/3$ (stellar scattering), 
			and on the right we plot a spectrum with $\kappa=7/3$ (gas-driven dynamics). 
			The top row shows a GW spectrum with $\Agw=10^{-15}$, while 
			the bottom row shows a GW spectrum with $\Agw=5\times10^{-15}$. 
			In all plots, $\gamma=13/3$, and the observing cadence is $c = 20 \: \mathrm{yr}^{-1}$. 
			The horizontal dotted lines indicate timing RMS of 
			$1 \; \mathrm{\mu}s$, $200 \; \mathrm{ns}$, and $100 \; \mathrm{ns}$.}
	\label{fig:timingRMS}
\end{figure}

Figure~\ref{fig:timingRMS} shows the white noise RMS values 
at which the GW power in the lowest frequency bin is equal to 
the white noise level, i.e., the transition from the weak-signal regime to the 
intermediate-signal regime, as a function of observation time. 
The left panel looks at GW spectra with $\kappa=10/3$, 
while the right panel looks at GW spectra with $\kappa=7/3$. 
We also compare two values of the GW amplitude, $\Agw = 10^{-15}$ (top row) and 
$\Agw = 5\times10^{-16}$ (bottom row). 
Within each panel, we compare the transition time for a power-law spectrum, 
a broken-power-law spectrum with $\fbend = 10^{-8} \: \mathrm{Hz}$, 
and a broken-power-law spectrum with 
$\fbend = 3\times10^{-9} \: \mathrm{Hz}$.

From the panels in Fig.~\ref{fig:timingRMS}, we see that if \fbend\ is smaller than or near \fstar, 
then the presence of a bend in the spectrum does not have a strong effect 
on the transition time, but if \fbend\ is much greater than \fstar, then the 
transition time becomes much larger. For example, for $\sigma = 1 \: \mu\mathrm{s}$, 
we transition to the intermediate-signal regime for a power-law spectrum with 
$A_\mathrm{gw}=10^{-15}$ after about 9 years of observation. 
For a broken-power-law spectrum with the same $A_\mathrm{gw}$ and $\kappa=10/3$ 
and $\fbend = 3\times10^{-9} \: \mathrm{Hz}$, the transition occurs after about 
10 years of observation. But if $\fbend = 10^{-8} \: \mathrm{Hz}$, we only reach 
the intermediate-signal regime after observing for more than 250 years -- 
in this case, for all practical purposes, NANOGrav will always be operating in the 
weak-signal regime. For a smaller $\kappa$, the bend in the GW spectrum is not as sharp, 
and so the bend does not delay the transition time as dramatically.

We expect environmental effects to have a significant impact on the evolution of 
SMBHBs in order to solve the last parsec problem. 
Whether these are detrimental to PTAs 
depends on which astrophysical mechanism is dominant 
and at what point the evolution of 
SMBHBs becomes GW-dominated. 
If the last parsec problem is solved by gas accretion, 
astrophysical interactions will not be seriously detrimental to PTAs. 
If the last parsec problem is solved by stellar scattering, the decrease in power at 
low frequencies could significantly hamper detection. However, this would require 
stellar densities that are significantly higher than what we expect based on observations 
of galaxies. The most likely scenario is that the decrease in power at low frequencies 
due to astrophysical effects will delay the transition to the intermediate-signal regime 
by no more than a few years.

\section{Scaling laws}
\label{sec:scalinglaws}

Here we derive analytic expressions for the SNR in the weak-signal and intermediate-signal regimes. \citet{2013CQGra..30v4015S} found that, for a power-law GW spectrum $P(f) \sim f^{-\gamma}$, in the weak-signal regime the SNR scales like $\SNR \sim T^{\gamma}$, while in the intermediate-signal regime the SNR scales like $\SNR \sim T^{1/2}$. We find similar results here, with the added complication that the spectrum can go like $f^{-\gamma}$ and $f^{-(\gamma-\kappa)}$ for different ranges of $f$. We break down our analysis in terms of the relationships between $\fstar$, $\fbend$, and $f_L$. For a given set of parameters, we will pass through up to three of these regimes as observation time increases.

\subsection{The weak-signal regime}

\begin{figure}[htb]
	\includegraphics[width=0.9 \columnwidth]{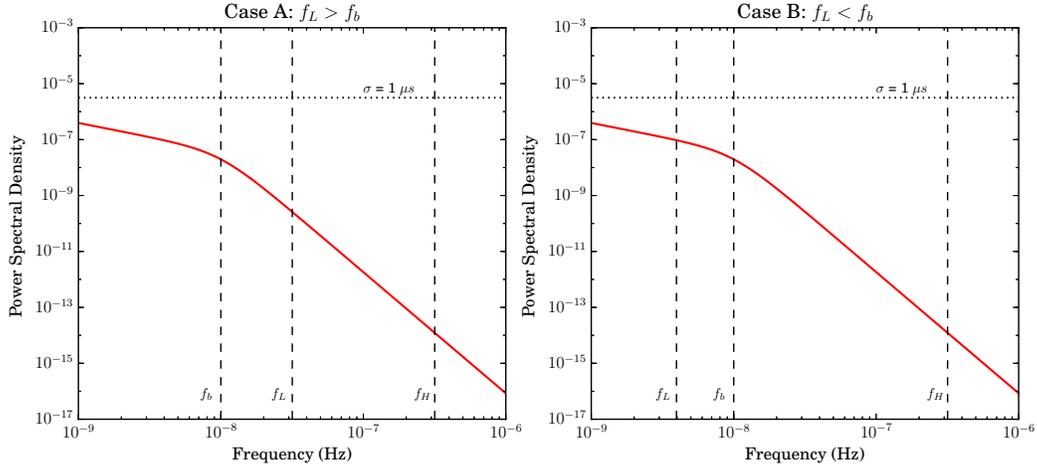}
	\caption{A schematic representation of the two possible scenarios in the weak-signal 
			regime when observing a broken-power-law spectrum. 
			Our PTA is sensitive to frequencies $f \in [f_L, f_H]$. The high-frequency 
			cutoff is set by the observing cadence, $f_H = c/2$ where $c=1/\Delta t$. The low-frequency 
			cutoff is set by the observation time, $f_L = 1/T$.
			In Case A, $f_L > \fbend$, and so we only observe the 
			steep portion of the spectrum. 
			In Case B, $f_L < \fbend$, which means we are sensitive to both the steep 
			and shallow portions of the spectrum.}
	\label{fig:weaksignal}
\end{figure}

In the weak-signal regime, the GW power spectral density is below the white noise for all frequencies $f \in [f_L, f_H]$, which means that we can simplify Eq.~\eqref{eq:SNR}:
\begin{eqnarray}
	\SNR \approx \HD \left[ 2T \int_{f_L}^{f_H} df \frac{\Pgw(f)^2}{\left(2\sigma^2 \Delta t \right)^2} \right]^{1/2} \,.
	\label{eq:SNR_weak}
\end{eqnarray}
We must consider two separate possibilities in the weak-signal regime: 
$f_L > \fbend$ and $f_L < \fbend$. 
If $f_L > \fbend$, then we are only integrating the spectrum in the steep-slope regime, 
and we can approximate the GW spectrum as $\Pgw(f) = bf^{-\gamma}$. 
If $f_L < \fbend$, then we must use the full expression for the broken-power-law spectrum. 
These two possibilities are shown in Fig.~\ref{fig:weaksignal} as Case A ($f_L > \fbend$) 
and Case B ($f_L < \fbend$).

First we consider Case A. Since $f_L < \fbend$, we can approximate the spectrum as 
$P(f) = bf^{-\gamma}$. When we substitute this into Eq.~\eqref{eq:SNR_weak}, 
we recover the scaling law for a power-law spectrum:
\begin{eqnarray}
	\SNR &\approx& \HD \left[ 2T \int_{f_L}^{f_H} df \frac{b^2 f^{-2\gamma}}{\left(2\sigma^2 \Delta t \right)^2} \right]^{1/2} \nonumber \\
		&\approx& \left(\sum_{IJ} \chi_{IJ}^2\right)^{1/2} \frac{b}{2\sigma^2 \Delta t} \frac{T^\gamma}{\sqrt{\gamma-1/2}} \,,
	\label{eq:SNR_weak1}
\end{eqnarray}
where we have used the fact that $f_L \ll f_H$ and $f_L = 1/T$ to simplify the result. 
Thus the SNR depends on the PTA configuration and stochastic background parameters according to
\begin{equation}
	\SNR \propto M \left(\frac{1}{\sigma^2 \Delta t}\right) b \, T^\gamma \,,
\end{equation}
where $M$ is the number of pulsars in the array.

For Case B, since $f_L < \fbend$, we must use the full expression for the broken-power-law spectrum:
\begin{equation}
	\SNR \approx \HD \frac{b}{2\sigma^2 \Delta t} \left\{ 2T \int_{f_L}^{f_H} df \frac{f^{-2\gamma}}{[1+(\fbend/f)^\kappa]^2} \right\}^{1/2} \,.
\end{equation}
This integral can be written in terms of an ordinary hypergeometric function:
\begin{equation}
	\SNR \approx \HD \frac{b}{2\sigma^2 \Delta t} \left\{ 2T \left[ \mathcal{I}(\gamma,\kappa,f_b,f_L) - \mathcal{I}(\gamma,\kappa,f_b,f_H) \right] \right\}^{1/2} \,,
\end{equation}
where $\mathcal{I}(\gamma,\kappa,f_b,f_1)$ is given by
\begin{equation}
	\mathcal{I}(\gamma,\kappa,f_b,f_1) = \frac{f_1^{-(2\gamma-1)}}{2\gamma-1} \F\left[ 2, \frac{2\gamma-1}{\kappa}, 1+\frac{2\gamma-1}{\kappa}, -\left(\frac{\fbend}{f_1}\right)^\kappa \right] \,.
\end{equation}
For an ordinary hypergeometric function $\F(a,b,c,z)$, if $z$ is small, it can be expanded as
\begin{equation}
	\F(a,b,c,z) = 1 + \frac{ab}{c} z + \mathcal{O}(z^2) \,.
\end{equation}
At $f_H$, $\fbend/f_H \ll 1$, and we can approximate the hypergeometric function as $1$. Then $\mathcal{I}(\gamma,\kappa,f_b,f_H)$ becomes simply
\begin{equation}
	\mathcal{I}(\gamma,\kappa,f_b,f_H) \approx \frac{f_H^{-(2\gamma-1)}}{2\gamma-1} \,.
\end{equation}
In order to do a similar expansion of $\mathcal{I}(\gamma,\kappa,\fbend,f_L)$, we must use identities to relate $\F(a,b,c,z)$ and $\F(a,b,c,z^{-1})$ [see Eq.~(15.8.2) of \cite{NHMF}]. Doing so allows us to approximate $\mathcal{I}(\gamma,\kappa,\fbend,f_L)$ as
\begin{eqnarray}
	\mathcal{I}(\gamma,\kappa,f_b,f_L) &\approx& \frac{f_L^{-(2\gamma-1)}}{2\gamma-1} \left\{ k_1(\gamma,\kappa) \left(\frac{f_L}{\fbend}\right)^{2\kappa} \left[ 1 - k_2(\gamma,\kappa) \left(\frac{f_L}{\fbend}\right)^{\kappa} \right] - k_3(\gamma,\kappa) \left(\frac{f_L}{\fbend}\right)^{2\gamma-1} \right\} \,,
\end{eqnarray}
where
\begin{eqnarray}
	k_1(\gamma,\kappa) &=& \frac{2\gamma-1}{2\gamma-2\kappa-1} \,, \\
	k_2(\gamma,\kappa) &=& \frac{4\gamma-4\kappa-2}{2\gamma-3\kappa-1} \,, \\
	k_3(\gamma,\kappa) &=& \frac{\pi (2\gamma-1) (2\gamma-\kappa-1)}{\kappa^2 \sin\left[\frac{\pi(2\gamma-1)}{\kappa}\right]} \,.
\end{eqnarray}
Putting this all together, and using the fact that $f_L/f_H \ll 1$, gives
\begin{widetext}
\begin{eqnarray}
	\SNR &\approx& \HD \frac{b}{2\sigma^2 \Delta t} \frac{T^\gamma}{\sqrt{\gamma-1/2}} \left\{ k_1(\gamma,\kappa) \left(\frac{f_L}{\fbend}\right)^{2\kappa} \left[ 1 - k_2(\gamma,\kappa) \left(\frac{f_L}{\fbend}\right)^{\kappa} \right] - k_3(\gamma,\kappa) \left(\frac{f_L}{\fbend}\right)^{2\gamma-1} \right\}^{1/2} \nonumber \\
		&\approx& \HD \frac{b}{2\sigma^2 \Delta t} \frac{T^\gamma}{\sqrt{\gamma-1/2}} \left\{ k_1(\gamma,\kappa) \: (\fbend T)^{-2\kappa} \left[ 1 - k_2(\gamma,\kappa) \left(\fbend T\right)^{-\kappa} \right] \right. \nonumber \\
			&& \left. \hspace{2in} - k_3(\gamma,\kappa) \: (\fbend T)^{-(2\gamma-1)} \right\}^{1/2} \,.
	\label{eq:SNR_weak2}
\end{eqnarray}
\end{widetext}
In the limit $T \gg \fbend^{-1}$, and the first term dominates, and the SNR can be approximated as:
\begin{equation}
	\SNR \approx \HD \frac{b}{2\sigma^2 \Delta t} \left[\frac{k_1(\gamma,\kappa)}{\gamma-1/2}\right]^{1/2} \fbend^{-\kappa} T^{\gamma-\kappa} \,.
\end{equation}
Thus the scaling law for the SNR in this case is
\begin{equation}
	\SNR \propto M \left(\frac{1}{\sigma^2 \Delta t}\right) b \, \fbend^{-\kappa} \, T^{\gamma-\kappa} \,.
\end{equation}

In summary, for frequencies below $\fbend$, the SNR increases with 
observation time as $T^\gamma$ in the large $T$ limit, whereas above $\fbend$ 
the SNR increases more slowly with observation time, as $T^{\gamma-\kappa}$. 
It is worth emphasizing that these scaling laws have different definitions 
of the large $T$ limit; in the case where $f_L > f_b$, ``large $T$'' means that $T$ must be 
large compared to $\Delta t$ (which is always true), whereas in the case where $f_L < f_b$, 
$T$ must be large compared to $\fbend^{-1}$. 
In both cases, the SNR is proportional to 
$1/(\sigma^2 \Delta t)$ -- which means that decreasing the residuals has a significant effect on the SNR -- 
and proportional to the number of pulsars in the array.

\subsection{Scaling laws: Intermediate-signal regime}

In the intermediate-signal regime, we cannot use the simplified version of \SNR\ 
that we made use of in Eq.~\eqref{eq:SNR_weak}. Instead we must use the full form of \SNR:
\begin{equation}
	\SNR = \HD \left\{ 2T \int_{f_L}^{f_H} df \frac{P_\mathrm{gw}(f)^2}{\left[P_\mathrm{gw}(f) + 2\sigma^2 \Delta t \right]^2} \right\}^{1/2} \,.
\end{equation}
In order to write this integral in terms of hypergeometric functions, 
we must use the piecewise form of $\Pgw(f)$ given in Eq.~\eqref{eq:Pgw_piecewise}. 
This will result in an overestimate of the SNR 
because the piecewise form of \Pgw\ is always above or equal to the exact form. 
However, we still get a good sense of how the SNR scales with the various parameters.

\begin{figure}[htb]
	\includegraphics[width=\textwidth]{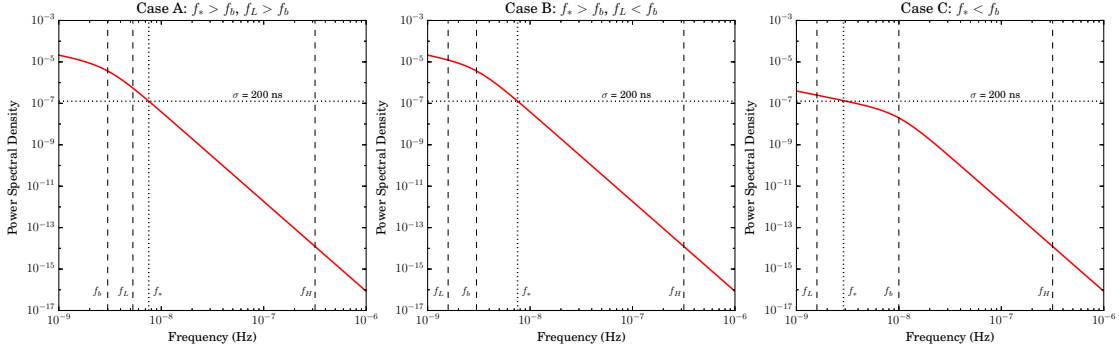}
	\caption{A schematic representation of the three possible scenarios in the 
			intermediate-signal regime when observing a broken-power-law spectrum. 
			Our PTA is sensitive to frequencies $f \in [f_L, f_H]$. The high-frequency 
			cutoff is set by the observing cadence, $f_H = c/2$ where $c=1/\Delta t$. The low-frequency 
			cutoff is set by the observation time, $f_L = 1/T$. 
			For Cases A and B, $\fstar > \fbend$, 
			so the GW signal intersects the white noise in the 
			steep portion of the spectrum. In Case A, the PTA is not yet sensitive to the 
			shallower portion of the spectrum (i.e., $f_L > \fbend$), 
			while in Case B the PTA is sensitive to both regions of the spectrum. 
			For Case C, $\fstar < \fbend$, 
			so only the shallow portion of the spectrum is above the white noise. 
			Assuming infinite observation time, all PTAs will end up in either Case B or Case C.}
	\label{fig:intsignal}
\end{figure}

There are three possible scenarios for the intermediate-signal regime 
based on the relationships between \fbend, \fstar, and $f_L$, 
which we illustrate Fig.~\ref{fig:intsignal}. 
First we consider the case where $\fstar > \fbend$, so that the GW spectrum intersects 
the white noise in the steep portion of the spectrum. If $f_L > \fbend$ (Case A in Fig.~\ref{fig:intsignal}), 
then we recover the result for a power-law spectrum in the intermediate-signal regime found by 
\citet{2015PhRvD..91d4048C}. The SNR can be written
\begin{eqnarray}
	\SNR &=& \HD \left\{ 2 T \int_{f_L}^{f_H} df \frac{b^2 f^{-2\gamma}}{[bf^{-\gamma} + 2\sigma^2 \Delta t]^2} \right\}^{1/2} \nonumber \\
		&=& \HD \left\{ 2T \left[ \mathcal{J} \left( \gamma, \frac{2\sigma^2 \Delta t}{b}, f_H \right) - \mathcal{J} \left( \gamma, \frac{2\sigma^2 \Delta t}{b}, f_L \right) \right] \right\}^{1/2} \,,
	\label{eq:SNR_int_J}
\end{eqnarray}
where
\begin{eqnarray}
	\mathcal{J}(\alpha,A,f_1) &\equiv& \int_0^{f_1} df \frac{1}{(1+Af^\alpha)^2} \nonumber \\
		&=& \frac{1}{\alpha} \frac{f_1}{1+Af_1^\alpha} + \frac{\alpha-1}{\alpha} f_1 \; \F\left(1,\frac{1}{\alpha},1+\frac{1}{\alpha},-Af_1^\alpha\right) \,.
	\label{eq:J_integral}
\end{eqnarray}
We simplify Eq.~\eqref{eq:SNR_int_J} as follows. 
At $f_L$, the white noise is much smaller than the GW signal, 
i.e., $2\sigma^2 \Delta t \ll b f_L^{-\gamma}$, 
which allows us to approximate the hypergeometric function evaluated at $f_L$ as 1. 
Then the second term in Eq.~\eqref{eq:SNR_int_J} can be approximated as
\begin{eqnarray}
	\mathcal{J} \left( \gamma, \frac{2\sigma^2 \Delta t}{b}, f_L \right) &\approx& \frac{1}{\gamma} f_L + \frac{\gamma-1}{\gamma} f_L \nonumber \\
		&\approx& \frac{1}{T} \,.
\end{eqnarray}
In order to simplify the first term in Eq.~\eqref{eq:integral_intermediate1} we must again use identities to relate $\F(a,b,c,z)$ to $\F(a,b,c,z^{-1})$ so that we can expand the hypergeometric function as a series [see Eq.~(15.8.2) of \cite{NHMF}]. Doing so gives
\begin{equation}
	\mathcal{J}\left(\gamma,\frac{2\sigma^2 \Delta t}{b},f_H\right) \approx \kappa(\gamma) \left( \frac{b}{2\sigma^2 \Delta t} \right)^{1/\gamma} \,,
\end{equation}
where
\begin{equation}
	\kappa(\alpha) \equiv \frac{\pi(\alpha-1)}{\alpha^2 \sin(\pi/\alpha)} \,.
	\label{eq:kappa}
\end{equation}
Putting this all together gives
\begin{equation}
	\SNR = \HD \left\{ 2T \left[ \kappa(\gamma) \left( \frac{b}{2\sigma^2 \Delta t} \right)^{1/\gamma} - \frac{1}{T} \right] \right\}^{1/2} \,.
	\label{eq:SNR_int}
\end{equation}

As observation time increases and $f_L$ decreases, we will eventually move into a regime where 
$f_L < \fbend$ (Case B in Fig.~\ref{fig:intsignal}). Then the SNR can be written as
\begin{equation}
	\SNR \approx \HD \left\{2T \int_{f_L}^{\fbend} df \frac{b^2 \fbend^{-2\kappa} f^{-2(\gamma-\kappa)}}{\left[b\fbend^{-\kappa} f^{-(\gamma-\kappa)}+2\sigma^2 \Delta t\right]^2} + 2T \int_{\fbend}^{f_H} df \frac{b^2 f^{-2\gamma}}{(bf^{-\gamma} + 2\sigma^2 \Delta t)^2} \right\}^{1/2} \,.
	\label{eq:SNR_int2_def}
\end{equation}
where we have used the piecewise form of the broken-power-law spectrum given in 
Eq.~\eqref{eq:Pgw_piecewise}. 
If $\gamma-\kappa > 1$, these integrals can be written in terms of hypergeometric functions:
\begin{eqnarray}
	\SNR &\approx& \HD \left\{ 2T \left[ \mathcal{J}\left(\gamma-\kappa,\frac{2\sigma^2 \Delta t}{b \fbend^{-\kappa}},\fbend\right) - \mathcal{J}\left(\gamma-\kappa,\frac{2\sigma^2 \Delta t}{b \fbend^{-\kappa}},f_L\right) \right] \right. \nonumber \\
		&& \left. + 2T \left[ \mathcal{J}\left(\gamma,\frac{2\sigma^2 \Delta t}{b},f_H\right) - \mathcal{J}\left(\gamma,\frac{2\sigma^2 \Delta t}{b},f_b\right) \right] \right\}^{1/2}
\end{eqnarray}
where $\mathcal{J}(\gamma,A,f_1)$ is defined in Eq.~\eqref{eq:J_integral}. We expand the hypergeometric functions as follows:
\begin{eqnarray}
	\mathcal{J}\left(\gamma-\kappa,\frac{2\sigma^2 \Delta t}{b \fbend^{-\kappa}},\fbend\right) &\approx& \fbend \left[ 1 - \frac{2}{\gamma-\kappa+1} \left(\frac{2\sigma^2 \Delta t}{b \fbend^{-\gamma}}\right) \right] \,, \\
	\mathcal{J}\left(\gamma-\kappa,\frac{2\sigma^2 \Delta t}{b \fbend^{-\kappa}},f_L\right) &\approx& \frac{1}{T} \,, \label{eq:integrals2} \\
	\mathcal{J}\left(\gamma,\frac{2\sigma^2 \Delta t}{b},f_H\right) &\approx& \kappa(\gamma) \left(\frac{b}{2\sigma^2 \Delta t}\right)^{1/\gamma} \,, \\
	\mathcal{J}\left(\gamma,\frac{2\sigma^2 \Delta t}{b},\fbend\right) &\approx& \fbend \left[ 1 - \frac{2}{\gamma+1} \left(\frac{2\sigma^2\Delta t}{b \fbend^{-\gamma}}\right) \right] \,. \label{eq:integrals4}
\end{eqnarray}
Putting this all together gives
\begin{equation}
	\SNR = \HD \left\{ 2T \left[ \kappa(\gamma) \left(\frac{b}{2\sigma^2\Delta t}\right)^{1/\gamma} - \frac{2\kappa}{(\gamma+1)(\gamma-\kappa+1)} \left(\frac{2\sigma^2\Delta t}{b \fbend^{-\gamma}}\right) \fbend - \frac{1}{T} \right] \right\}^{1/2} \,.
	\label{eq:SNR_int2}
\end{equation}
The middle term in Eq.~\eqref{eq:SNR_int2} may be negligible depending on the relative power in the 
GW spectrum compared to the white noise at \fbend. We include it here to differentiate between 
this case and Case A; however, for all practical purposes there is very little difference in the SNR 
between these two cases. 
In the special case where $\gamma-\kappa=1$ (which is the case for $\kappa=10/3$), 
the first integral in Eq.~\eqref{eq:SNR_int2_def} can be done exactly: 
\begin{eqnarray}
	\int_{f_L}^{\fbend} df \frac{b^2 \fbend^{-2(\gamma-1)} f^{-2}}{\left[b\fbend^{-(\gamma-1)} f^{-1}+2\sigma^2 \Delta t\right]^2} &=& \frac{\fbend-f_L}{\left( 1 + \frac{2\sigma^2 \Delta t}{b \fbend^{-\kappa}} \fbend \right) \left( 1 + \frac{2\sigma^2 \Delta t}{b \fbend^{-\kappa}} f_L \right)} \nonumber \\
		&\approx& \fbend \left( 1 - \frac{2\sigma^2 \Delta t}{b \fbend^{-\gamma}} \right) - \frac{1}{T} \,,
\end{eqnarray}
where in going from the first line to the second, we have used the fact that $f_L \ll \fbend$ and $2\sigma^2 \Delta t \ll b\fbend^{-\gamma}$. 
Then the SNR becomes 
\begin{equation}
	\SNR = \HD \left\{ 2T \left[ \kappa(\gamma) \left(\frac{b}{2\sigma^2\Delta t}\right)^{1/\gamma} - \frac{\gamma-1}{\gamma+1} \left(\frac{2\sigma^2 \Delta t}{b \fbend^{-\gamma}}\right) \fbend - \frac{1}{T} \right] \right\}^{1/2} \,.
	\label{eq:SNR_int2_special}
\end{equation}

Now we consider the case where $\fstar < \fbend$ (Case C in Fig.~\ref{fig:intsignal}), so that only the flatter part of the spectrum is above the white noise. Then the SNR can be approximated as
\begin{equation}
	\SNR = \HD \left\{ 2T \int_{f_L}^{\fbend} df \frac{b^2 \fbend^{-2\kappa} f^{-2(\gamma-\kappa)}}{\left[b\fbend^{-\kappa} f^{-(\gamma-\kappa)}+2\sigma^2 \Delta t\right]^2} + 2T \int_{\fbend}^{f_H} df \frac{b^2 f^{-2\gamma}}{(2\sigma^2 \Delta t)^2} \right\}^{1/2} \,.
	\label{eq:SNR_int_case1}
\end{equation}
The second integral is the same one we did in Eq.~\eqref{eq:SNR_weak1}, except the lower bound 
on the integral is \fbend\ instead of $f_L$. 
If $\gamma-\kappa > 1$, 
the first integral can be expressed in terms of hypergeometric functions, 
\begin{equation}
	\int_{f_L}^{\fbend} df \frac{b^2 \fbend^{-2\kappa} f^{-2(\gamma-\kappa)}}{\left[ b\fbend^{-\kappa} f^{-(\gamma-\kappa)}+2\sigma^2 \Delta t \right]^2} = \mathcal{J}\left(\gamma-\kappa,\frac{2\sigma^2 \Delta t}{bf_b^{-\kappa}},f_b\right) - \mathcal{J}\left(\gamma-\kappa,\frac{2\sigma^2 \Delta t}{bf_b^{-\kappa}},f_L\right) \,,
	\label{eq:integral_intermediate1}
\end{equation}
where $\mathcal{J}$ is defined in Eq.~\eqref{eq:J_integral}. 
At $f=f_L$, the white noise is much lower than the GW spectrum, i.e., $2\sigma^2\Delta t \ll b \fbend^{-\kappa} f_L^{-(\gamma-\kappa)}$. Therefore we can approximate the hypergeometric function as 1, and the second term in Eq.~\eqref{eq:integral_intermediate1} can be approximated as
\begin{eqnarray}
	\mathcal{J}\left(\gamma-\kappa,\frac{2\sigma^2 \Delta t}{bf_b^{-\kappa}},f_L\right) &\approx& \frac{f_L}{\gamma-\kappa} + f_L \left(1-\frac{1}{\gamma-\kappa}\right) \,, \nonumber \\
		&\approx& \frac{1}{T} \,.
\end{eqnarray}
In order to simplify the first term in Eq.~\eqref{eq:integral_intermediate1} we must again use identities to relate $\F(a,b,c,z)$ to $\F(a,b,c,z^{-1})$ so that we can expand the hypergeometric function as a series. Doing so gives
\begin{eqnarray}
	\mathcal{J}\left(\gamma-\kappa,\frac{2\sigma^2 \Delta t}{bf_b^{-\kappa}},\fbend\right) &\approx& \kappa(\gamma-\kappa) \left( \frac{b \fbend^{-\kappa}}{2\sigma^2 \Delta t} \right)^{1/(\gamma-\kappa)} \,,
\end{eqnarray}
where $\kappa$ is defined in Eq.~\eqref{eq:kappa}. 
Putting this all together gives
\begin{eqnarray}
	\SNR &=& \HD \left\{ 2T \left[ \kappa(\gamma-\kappa) \left(\frac{b \fbend^{-\kappa}}{2\sigma^2 \Delta t}\right)^{1/(\gamma-\kappa)}  - \frac{1}{T} \right] + \frac{T}{\gamma-1/2} \left(\frac{b}{2\sigma^2\Delta t}\right)^2 \fbend^{-2\gamma+1} \right\}^{1/2} \,,
	\label{eq:SNR_int1}
\end{eqnarray}
In the special case where $\gamma-\kappa=1$ (as is the case for $\kappa = 10/3$), 
we can do the first integral in Eq.~\eqref{eq:SNR_int_case1} exactly:
\begin{eqnarray}
	\int_{f_L}^{\fbend} df \frac{b^2 \fbend^{-2(\gamma-1)} f^{-2}}{\left[b\fbend^{-(\gamma-1)} f^{-1}+2\sigma^2 \Delta t\right]^2} &=& \frac{\fbend-f_L}{\left( 1 + \frac{2\sigma^2 \Delta t}{b \fbend^{-\kappa}} \fbend \right) \left( 1 + \frac{2\sigma^2 \Delta t}{b \fbend^{-\kappa}} f_L \right)} \nonumber \\
		&\approx& \left(\frac{b}{2\sigma^2 \Delta t}\right) \fbend^{-\gamma+1} - \frac{1}{T} \,.
\end{eqnarray}
Substituting this into Eq.~\eqref{eq:SNR_int_case1} gives
\begin{eqnarray}
	\SNR &=& \HD \left\{ 2T \left[ \left(\frac{b}{2\sigma^2 \Delta t}\right) \fbend^{-\gamma+1} - \frac{1}{T} \right] + \frac{T}{\gamma-1/2} \left( \frac{b}{2\sigma^2\Delta t} \right)^2 \fbend^{-2\gamma+1} \right\}^{1/2} \,,
	\label{eq:SNR_int1_special}
\end{eqnarray}

All of the intermediate-signal regime scaling laws are remarkably similar. 
In the large $T$ limit, we can neglect the $1/T$ term, and 
so all forms of the SNR scale as $\SNR \sim T^{1/2}$. They differ in their 
scaling with GW spectrum amplitude $b$, timing RMS $\sigma$, and the observing 
cadence $c = 1/\Delta t$, depending on whether the transition to the intermediate-signal 
regime is in the shallow portion of the spectrum or the steep portion of the spectrum. 
If it transition to the intermediate-signal regime in the steep portion of the spectrum, the 
SNR is insensitive to the residuals and the observing cadence, whereas if it transitions 
in the shallow part of the spectrum, there is a somewhat stronger dependence. 
In all of these cases, the SNR is most sensitive to the number of pulsars in the array, 
which enters into the SNR 
through the sum of the squares of the Hellings-Downs coefficients.

\section{Numerical comparisons}
\label{sec:numerical}
In this section we compare the analytic forms for $\SNR$ derived in Sec.~\ref{sec:scalinglaws} 
to numerical calculations of $\SNR$ performed in the time-domain for several combinations of 
the parameters $\kappa$ and $\fbend$. Our goal is to demonstrate how well these 
scaling relationships approximate the SNR, plus show how the detector moves into the 
different regimes depending on the GW spectrum parameters and properties of our PTA. 
In all of these cases, we consider a PTA consisting of 20 pulsars observed with a 
cadence of $c=20 \; \mathrm{yr}^{-1}$.

\begin{figure}[htb]
	\begin{center}
	\includegraphics[width=0.47 \columnwidth]{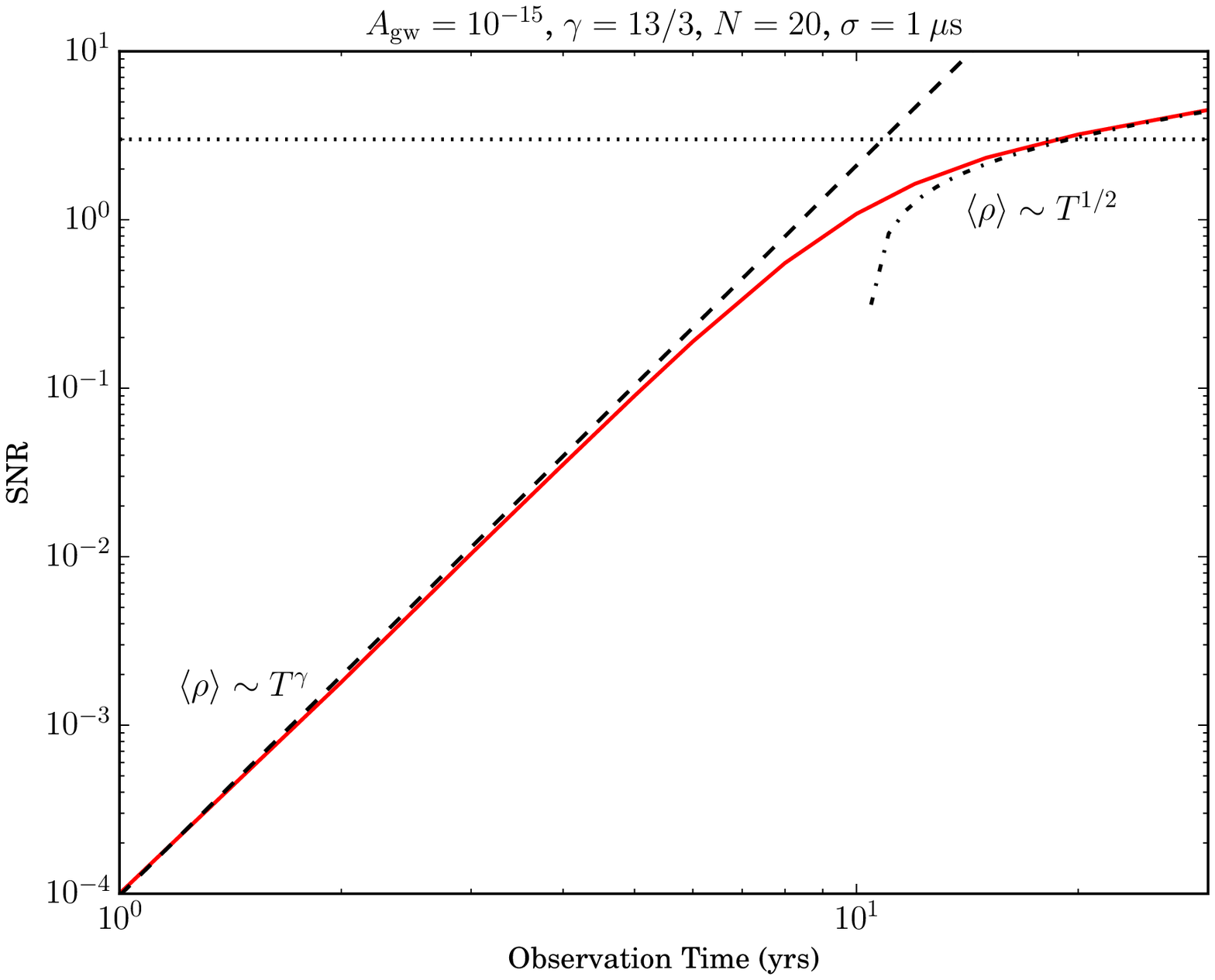}
	\includegraphics[width=0.47 \columnwidth]{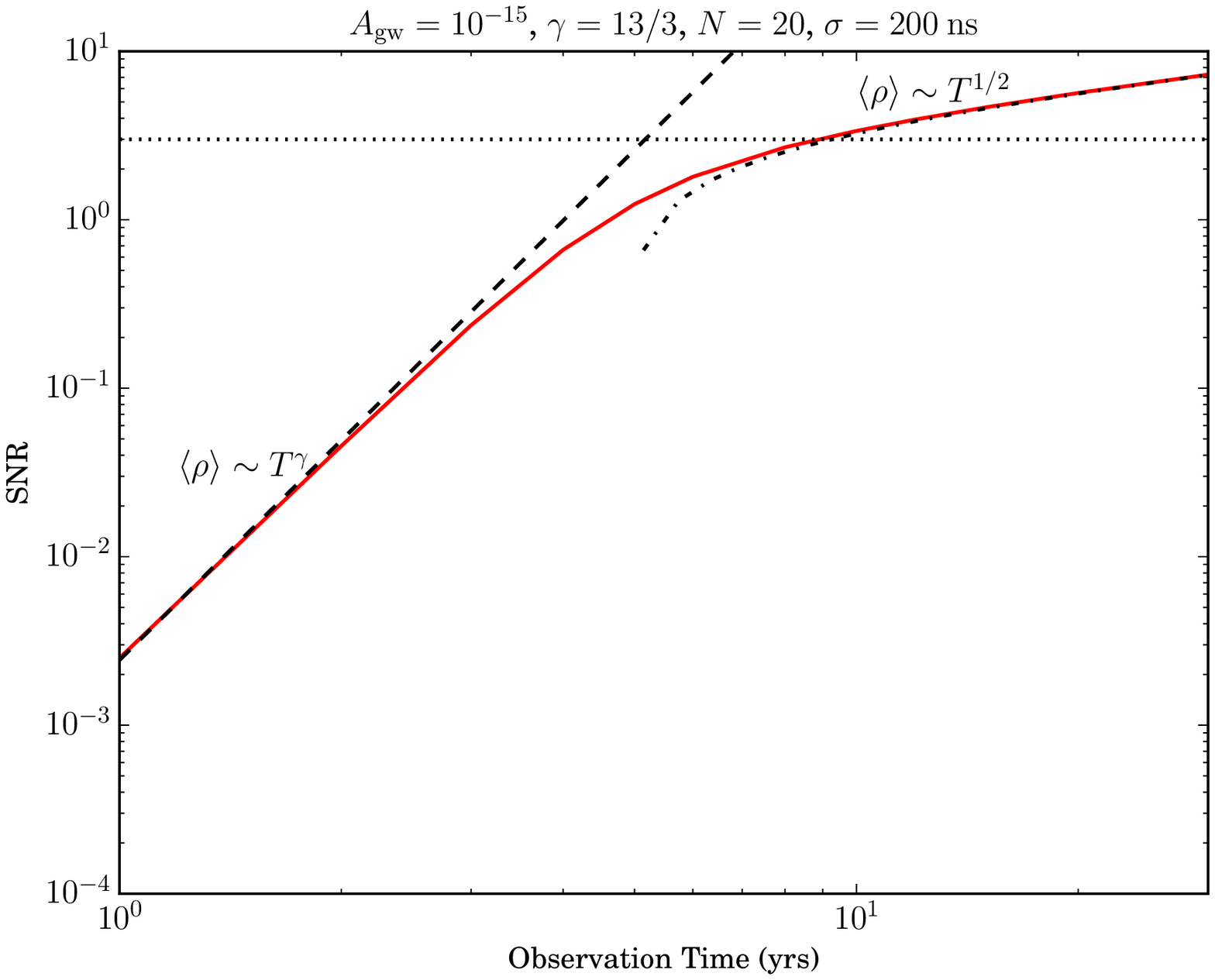}
	\caption{The SNR of a PTA consisting of 20 pulsars 
			observed with a cadence of $c=20 \; \mathrm{yr}^{-1}$. 
			The left panel shows an array with timing RMS of $1 \: \mu\mathrm{s}$, 
			and the right shows an array with timing RMS of $200 \: \mathrm{ns}$. 
			The stochastic background is a power law 
			with $A_\mathrm{gw} = 10^{-15}$ and $\gamma=13/3$. 
			The solid red line is the result of computing the S/N in the time domain. 
			The dashed line shows the weak-signal regime scaling law given in 
			Eq.~\eqref{eq:SNR_weak1}, and the dashed-dotted line shows the 
			intermediate-signal regime  
			scaling law given in Eq.~\eqref{eq:SNR_int}. At early times the SNR scales 
			as $T^\gamma$, while at late times the SNR scales as $T^{1/2}$. 
			The dotted horizontal line indicates an SNR of 3. 
			The array with $\sigma=1 \: \mu\mathrm{s}$ reaches an SNR of 3 after about 
			19 years, while the array with $\sigma=200 \: \mathrm{ns}$ reaches it 
			after about 9 years.}
	\label{fig:scaling_nobend}
	\end{center}
\end{figure}

First we review the scaling laws for a power-law GW spectrum with $\Agw=10^{-15}$ 
and $\gamma=13/3$. Figure~\ref{fig:scaling_nobend} shows the SNR for a PTA with 
$\sigma = 1 \: \mu\mathrm{s}$ (left panel) and $\sigma = 200 \: \mathrm{ns}$ (right panel). 
Both arrays begin in the weak-signal regime, 
where the SNR scales as $T^\gamma$ [Eq.~\eqref{eq:SNR_weak1}], then transition to the 
intermediate-signal regime, where the SNR scales as $T^{1/2}$ [Eq.~\eqref{eq:SNR_int}]. 
Assuming a detection threshold of $\SNR = 3$, the GW signal becomes detectable after 
about 19 years of observation for the array with $\sigma = 1 \: \mu\mathrm{s}$, 
and about 9 years of observation for the array with $\sigma = 200 \: \mathrm{ns}$.

\begin{figure}[htb]
	\begin{center}
	\includegraphics[width=0.47 \columnwidth]{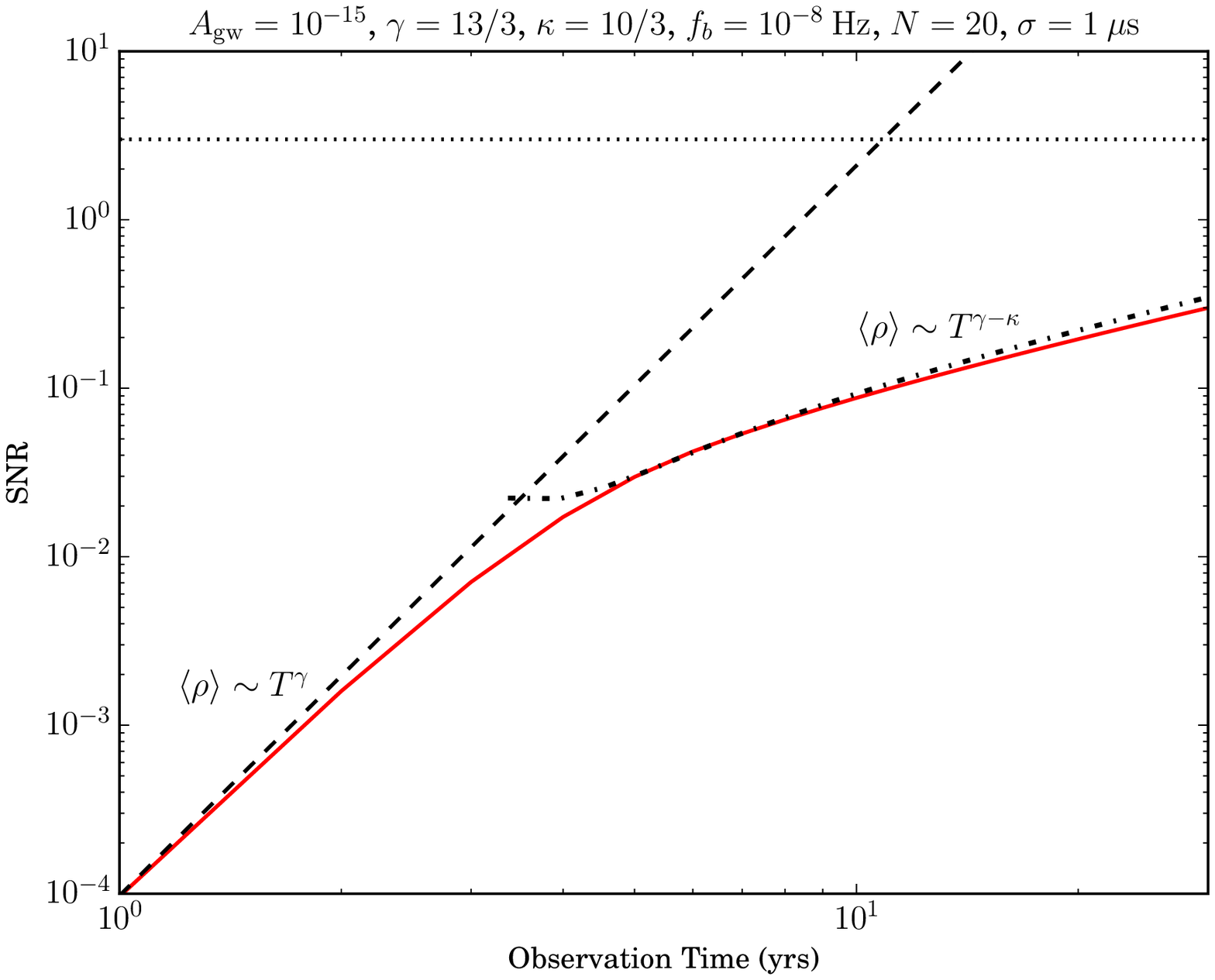}
	\includegraphics[width=0.47 \columnwidth]{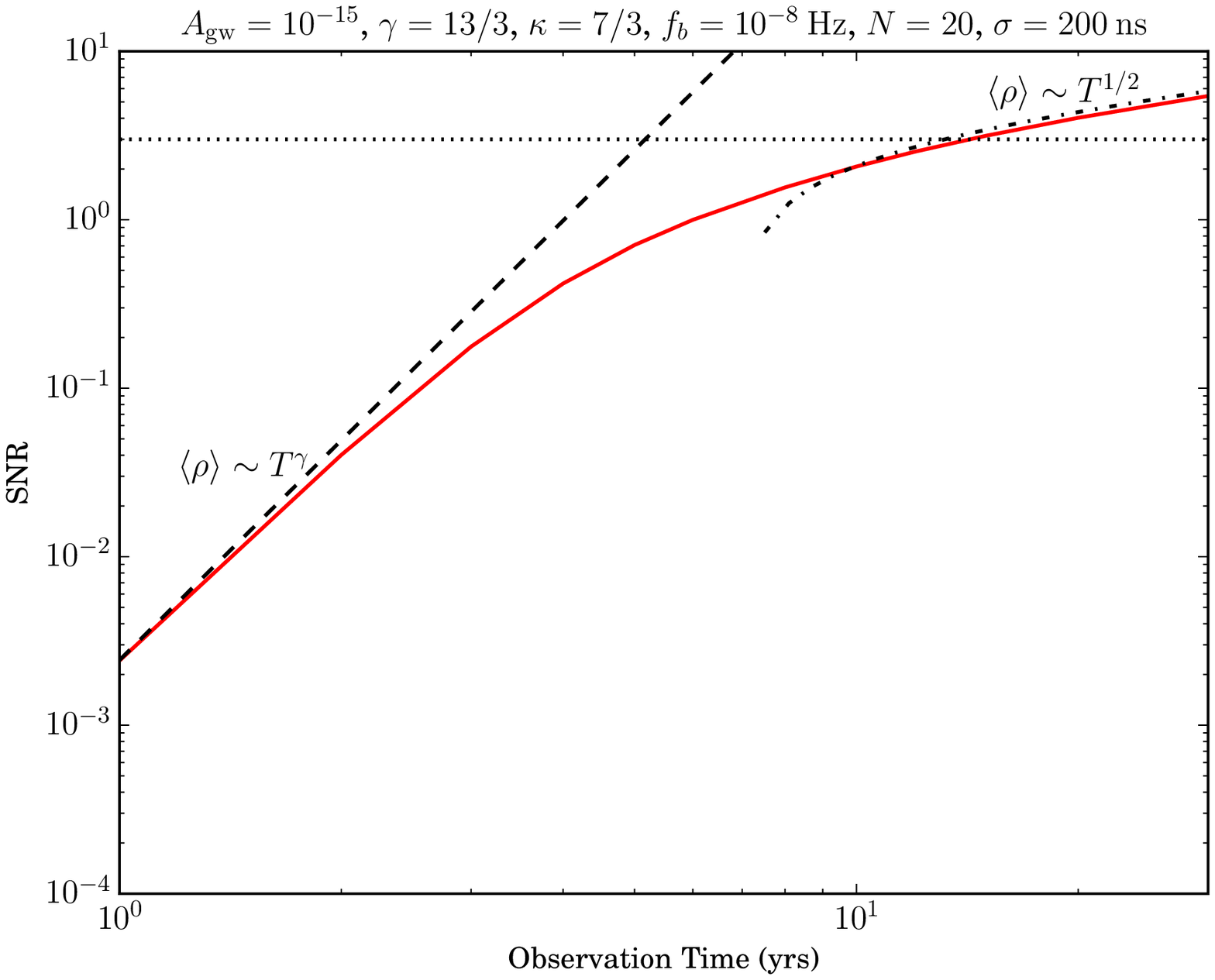}
	\caption{The SNR of a PTA consisting of 20 pulsars 
			observed with a cadence of $c=20 \; \mathrm{yr}^{-1}$ 
			with a timing RMS of $1 \; \mu\mathrm{s}$ (left) and $200 \: \mathrm{ns}$ (right). 
			The GW background has $A_\mathrm{gw} = 10^{-15}$, $\gamma=13/3$, 
			and $\fbend = 10^{-8} \; \mathrm{Hz}$. 
			In the left panel $\kappa=10/3$, and in the right $\kappa=7/3$. 
			The solid red line is the result of computing the S/N in the time domain. 
			The dotted horizontal line indicates an SNR of 3. 
			At early times the SNR of the PTA with $\sigma = 1 \: \mu\mathrm{s}$ 
			is described by the weak-signal scaling law in Eq.~\eqref{eq:SNR_weak1} 
			and scales like $T^\gamma$, as shown by the dashed line. 
			Once it is sensitive to the turnover in the spectrum, 
			it is described by Eq.~\eqref{eq:SNR_weak2} and scales like $T^{\gamma-\kappa}$, 
			indicated by the dashed-dotted line. This PTA does not enter the intermediate-signal 
			regime until more than 250 years of observation, 
			and the stochastic background is undetectable. 
			The PTA with $\sigma = 200 \: \mathrm{ns}$ begins in the weak-signal regime 
			[Eq.~\eqref{eq:SNR_weak1}; dashed line]. The turnover in the spectrum delays 
			the transition into the intermediate-signal regime until after about 6 years of observation, 
			after which the SNR scales like $T^{1/2}$ [Eq.~\eqref{eq:SNR_int1}; dashed-dotted line]. 
			The stochastic background is detectable after about 14 years, compared to 
			9 years for the the same PTA configuration observing a spectrum with no bend.}
	\label{fig:scaling1}
	\end{center}
\end{figure}

Now we consider a GW spectrum with $\Agw=10^{-15}$, $\gamma=13/3$, 
and $\fbend = 10^{-8} \; \mathrm{Hz}$. 
Since this spectrum turns over at high frequency, the SNR grows significantly more slowly 
than we saw in the previous examples. 
Figure~\ref{fig:scaling1} shows the SNR computed numerically along with the appropriate scaling laws 
for an array with $\sigma = 1 \: \mu\mathrm{s}$ (left) and $\sigma = 200 \: \mathrm{ns}$ (right). 
We also compare a turnover with $\kappa=10/3$ (left) and $\kappa=7/3$ (right). 
Initially the SNR grows as $T^\gamma$ [Eq.~\eqref{eq:SNR_weak1}], 
then turns over after about 3 years of observation time and begins growing more slowly 
as $T^{\gamma-\kappa}$ [Eq.~\eqref{eq:SNR_weak2}]. 
The PTA with $\sigma = 1 \: \mu\mathrm{s}$ does not enter the intermediate-signal regime until 
after more than 250 years of observation. 
For this configuration, the SNR is so low that the GW signal is not detectable. 
For the PTA with $\sigma = 200 \: \mathrm{ns}$, the SNR is much higher when the 
spectrum turns over and the turnover is not as sharp, 
and so the turnover only delays detection by about five years. The PTA 
reaches the intermediate-signal regime after about 6 years of observation (compared to 5 years 
for a GW spectrum with no turnover), and reaches our detection threshold of $\SNR=3$ after about 
14 years.

\begin{figure}[htb]
	\includegraphics[width=0.47 \columnwidth]{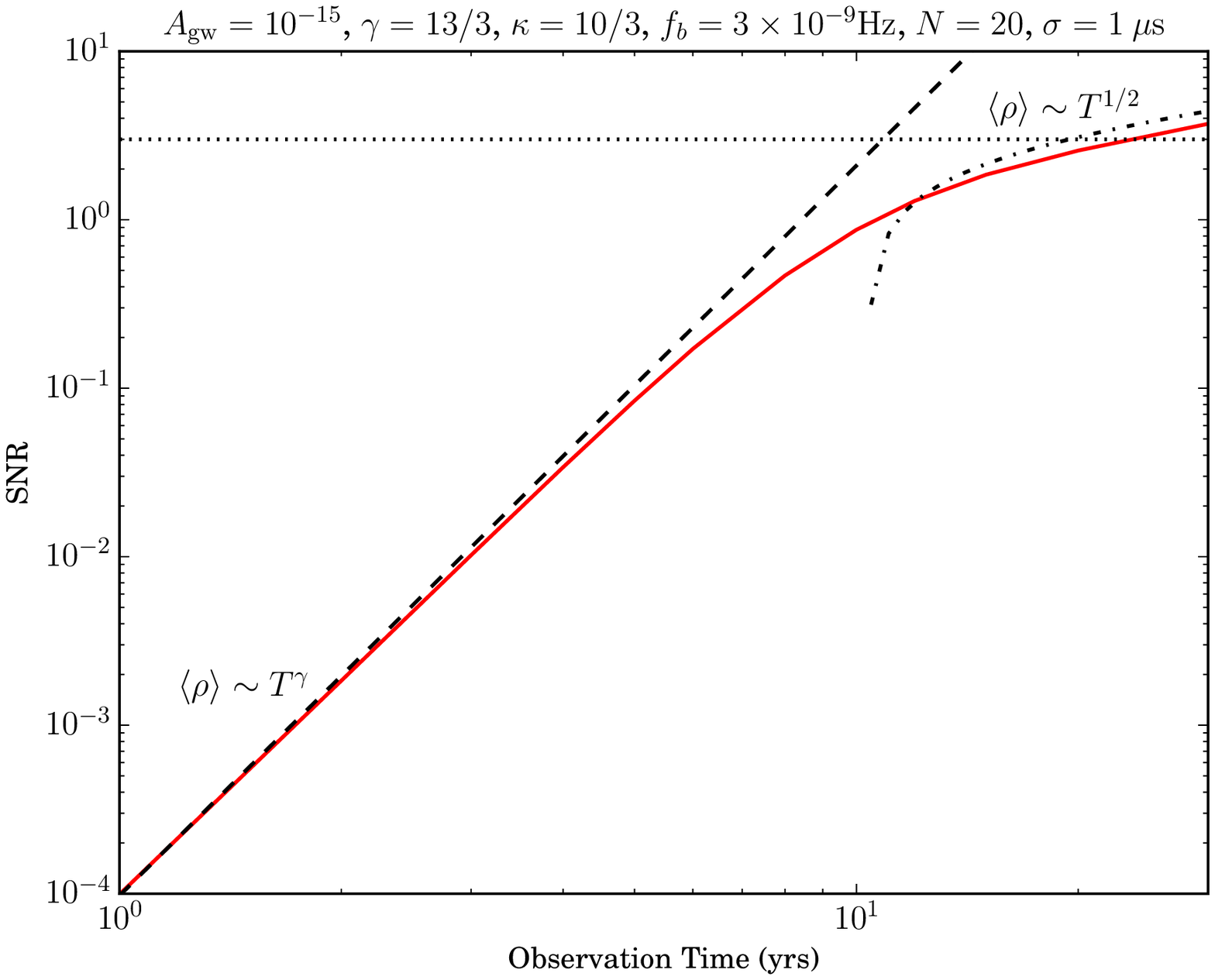}
	\includegraphics[width=0.47 \columnwidth]{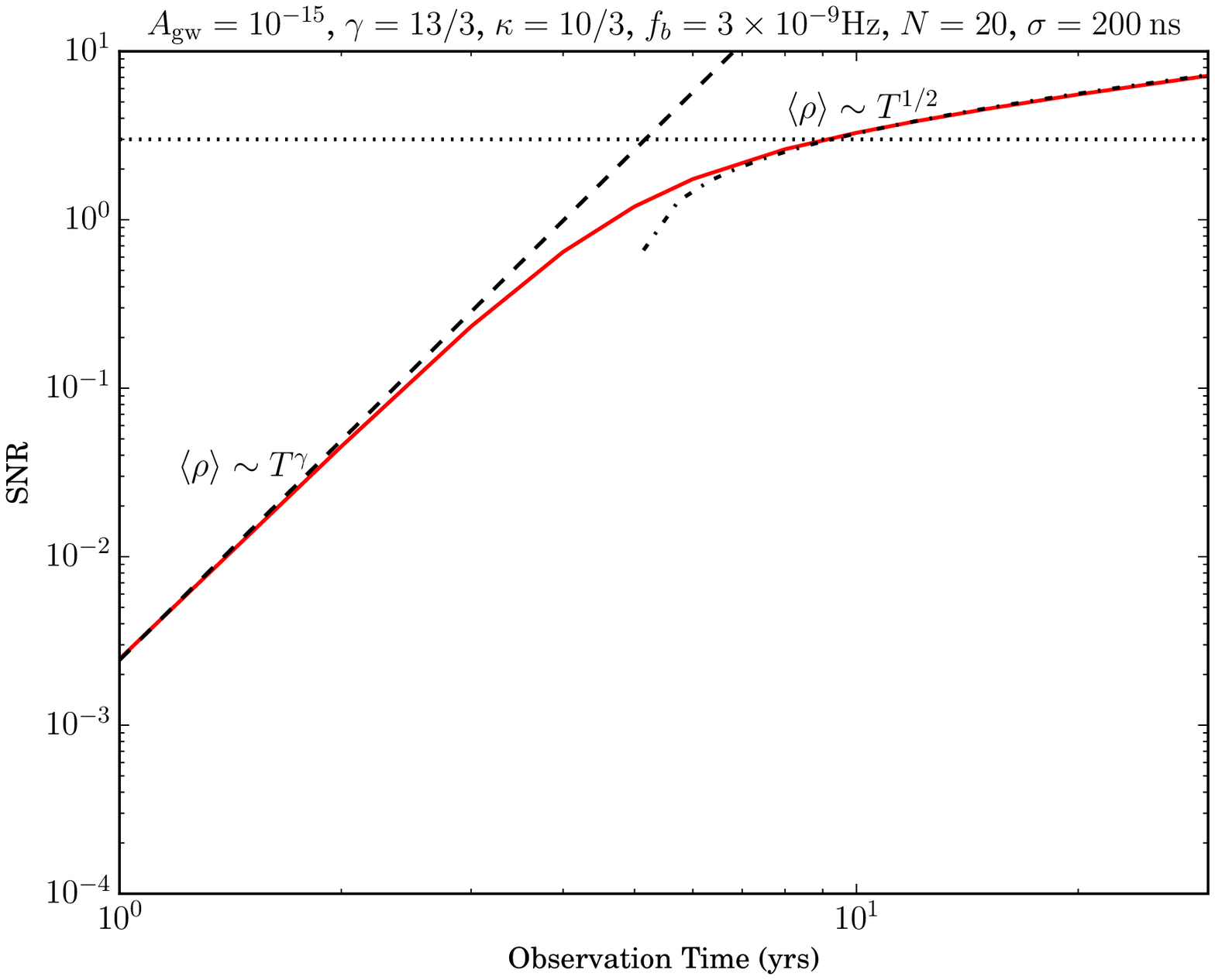}
	\caption{The SNR of a PTA consisting of 20 pulsars 
			observed with a cadence of $c=20 \; \mathrm{yr}^{-1}$ 
			with timing RMS of $1 \; \mu\mathrm{s}$ (left) and $200 \: \mathrm{ns}$ (right). 
			The GW background has $A_\mathrm{gw} = 10^{-15}$, 
			$\gamma=13/3$, $\kappa=10/3$, and $\fbend = 3 \times 10^{-9} \; \mathrm{Hz}$. 
			The solid red line is the result of computing the S/N in the time domain. 
			The dashed line shows the weak-field scaling law given in Eq.~\eqref{eq:SNR_weak1}, 
			and the dashed-dotted line shows the intermediate-field scaling law given in 
			Eq.~\eqref{eq:SNR_int2_special}. 
			At early times the SNR scales as $T^\gamma$, 
			while for late times the SNR scales as $T^{1/2}$. Since the turnover in the spectrum is 
			at low frequency, there is very little difference between these cases and observing a 
			power-law spectrum (Fig.~\ref{fig:scaling_nobend}). 
			For the PTA with $\sigma = 1 \: \mu\mathrm{s}$, detection is delayed by a few years; 
			for the PTA with $\sigma = 200 \: \mathrm{ns}$, detection is delayed by less than a year.}
	\label{fig:scaling3}
\end{figure}

Finally, we consider a GW spectrum with 
$\fbend = 3 \times 10^{-9} \; \mathrm{Hz}$ and $\kappa=10/3$, 
and a PTA with timing RMS of $\sigma = 1 \; \mu\mathrm{s}$. 
The SNR and scaling laws are plotted in Fig.~\ref{fig:scaling3}. 
This GW spectrum turns over after about 10 years of observation time, 
at which point the PTA has already transitioned to the intermediate-signal regime. 
The SNR scales as $T^\gamma$ [Eq.~\eqref{eq:SNR_weak1}] at early times, 
and then at late times it scales as $T^{1/2}$ [Eq.~\eqref{eq:SNR_int2_special}]. 
There is very little difference between the SNR for this GW spectrum 
and the SNR for a power-law GW spectrum since the spectrum does not turn over 
until after the PTA reaches the intermediate-signal regime.

\section{Conclusions}
\label{sec:conclusions}

In this paper we extended the work of \citet{2013CQGra..30v4015S} and \citet{2015PhRvD..91d4048C} 
by using the optimal cross-correlation statistic to derive scaling laws for the SNR 
for a GW stochastic background that includes environmental effects. 
These effects cause the spectrum to turn over at some frequency \fbend, 
depleting the strain spectrum at low frequencies.  
We found that this can have a significant effect on the SNR depending on 
the PTA configuration and when the spectrum turns over. 
If the spectrum turns over after we enter the intermediate-signal regime, 
the bend will have little effect on the SNR. 
However, if the spectrum turns over before we enter the intermediate-signal regime, 
the SNR will be significantly lower compared to observing a power-law spectrum. 
This is because once the spectrum turns over, the SNR grows much more slowly with observation time. 
Initially, $\SNR \sim T^\gamma$, but if the spectrum turns over in the weak-signal regime, 
the SNR will also turn over and scale as $\SNR \sim T^{\gamma-\kappa}$. In the 
intermediate-signal regime, $\SNR \sim T^{1/2}$ regardless of the shape of the spectrum, 
but the bend has some effect on how the SNR scales with observing cadence and timing RMS.

Here we focused on two environmental effects: 
stellar scattering, which corresponds to a turnover coefficient of $\kappa=10/3$, 
and gas-driven dynamics, which corresponds to a turnover coefficient of $\kappa=7/3$. 
We also considered two representative potential values of the turnover frequency, 
$\fbend=10^{-8} \: \mathrm{Hz}$ and $\fbend = 3\times10^{-9} \: \mathrm{Hz}$. 
If stellar scattering is the dominant mechanism at low frequencies, the resulting 
decrease in power may significantly hamper efforts to observe the stochastic background 
if the turnover is relatively high. However, stellar densities around SMBHs are unlikely to be 
high enough to produce such a high-frequency turnover. 
For gas-driven dynamics, the turnover in the spectrum is not as sharp, 
and therefore the effect on the SNR is not as significant regardless of the turnover frequency. 
Therefore we conclude that environmental effects are unlikely to impede 
detection of the stochastic background; however, they may delay detection by a few years.

In the future we plan to generate realistic spectra using numerical codes 
in order to better estimate how the time to detection depends on astrophysical and 
cosmological properties. 
This includes varying the GW spectrum amplitude \Agw, 
which we have taken to be $\Agw = 10^{-15}$ 
in the simulations in Sec.~\ref{sec:numerical}. 
This is near the current limit placed by PTAs and coincides with models by 
\citet{2013MNRAS.433L...1S} and \citet{2015MNRAS.447.2772R}; 
however, recent work suggests the actual background may be a factor of 2 or 3 smaller 
\citep{2016arXiv160607484R, 2016MNRAS.463L...6S, 2016arXiv160601900K}.  
A decrease in \Agw\ will have a significant effect on the detection of GWs, as it will 
decrease the SNR and increase 
the amount of observation time needed to enter the intermediate-signal regime. 
We also plan to include orbital eccentricity, which 
depletes the stochastic background strain at low frequencies in a different way than 
the environmental effects considered here.

\acknowledgements
This work was supported through the 
National Science Foundation (NSF) PIRE
program award number 0968296. 
We thank Justin Ellis, Sean McWilliams, Chiara Mingarelli, Joseph Simon, 
and Steve Taylor for helpful comments.

\bibliographystyle{apsrev} 
\bibliography{master}

\newcommand{\mnras}{Mon.~Not.~R.~Astron.~Soc.}\newcommand{\apjl}{Astrophys.~J.,~Lett.}
\begin{thebibliography}{27}
\expandafter\ifx\csname natexlab\endcsname\relax\def\natexlab#1{#1}\fi
\expandafter\ifx\csname bibnamefont\endcsname\relax
  \def\bibnamefont#1{#1}\fi
\expandafter\ifx\csname bibfnamefont\endcsname\relax
  \def\bibfnamefont#1{#1}\fi
\expandafter\ifx\csname citenamefont\endcsname\relax
  \def\citenamefont#1{#1}\fi
\expandafter\ifx\csname url\endcsname\relax
  \def\url#1{\texttt{#1}}\fi
\expandafter\ifx\csname urlprefix\endcsname\relax\def\urlprefix{URL }\fi
\providecommand{\bibinfo}[2]{#2}
\providecommand{\eprint}[2][]{\url{#2}}

\bibitem[{\citenamefont{{Taylor} et~al.}(2016)\citenamefont{{Taylor},
  {Vallisneri}, {Ellis}, {Mingarelli}, {Lazio}, and {van
  Haasteren}}}]{2016ApJ...819L...6T}
\bibinfo{author}{\bibfnamefont{S.~R.} \bibnamefont{{Taylor}}},
  \bibinfo{author}{\bibfnamefont{M.}~\bibnamefont{{Vallisneri}}},
  \bibinfo{author}{\bibfnamefont{J.~A.} \bibnamefont{{Ellis}}},
  \bibinfo{author}{\bibfnamefont{C.~M.~F.} \bibnamefont{{Mingarelli}}},
  \bibinfo{author}{\bibfnamefont{T.~J.~W.} \bibnamefont{{Lazio}}},
  \bibnamefont{and} \bibinfo{author}{\bibfnamefont{R.}~\bibnamefont{{van
  Haasteren}}}, \bibinfo{journal}{\apjl} \textbf{\bibinfo{volume}{819}},
  \bibinfo{eid}{L6} (\bibinfo{year}{2016}), \eprint{1511.05564}.

\bibitem[{\citenamefont{{Rajagopal} and {Romani}}(1995)}]{1995ApJ...446..543R}
\bibinfo{author}{\bibfnamefont{M.}~\bibnamefont{{Rajagopal}}} \bibnamefont{and}
  \bibinfo{author}{\bibfnamefont{R.~W.} \bibnamefont{{Romani}}},
  \bibinfo{journal}{\apj} \textbf{\bibinfo{volume}{446}}, \bibinfo{pages}{543}
  (\bibinfo{year}{1995}), \eprint{astro-ph/9412038}.

\bibitem[{\citenamefont{{Jaffe} and {Backer}}(2003)}]{2003ApJ...583..616J}
\bibinfo{author}{\bibfnamefont{A.~H.} \bibnamefont{{Jaffe}}} \bibnamefont{and}
  \bibinfo{author}{\bibfnamefont{D.~C.} \bibnamefont{{Backer}}},
  \bibinfo{journal}{\apj} \textbf{\bibinfo{volume}{583}}, \bibinfo{pages}{616}
  (\bibinfo{year}{2003}), \eprint{astro-ph/0210148}.

\bibitem[{\citenamefont{{Wyithe} and {Loeb}}(2003)}]{2003ApJ...590..691W}
\bibinfo{author}{\bibfnamefont{J.~S.~B.} \bibnamefont{{Wyithe}}}
  \bibnamefont{and} \bibinfo{author}{\bibfnamefont{A.}~\bibnamefont{{Loeb}}},
  \bibinfo{journal}{\apj} \textbf{\bibinfo{volume}{590}}, \bibinfo{pages}{691}
  (\bibinfo{year}{2003}), \eprint{astro-ph/0211556}.

\bibitem[{\citenamefont{{Z.~Arzoumanian et al.~(The NANOGrav
  Collaboration)}}(2016)}]{2016ApJ...821...13A}
\bibinfo{author}{\bibnamefont{{Z.~Arzoumanian et al.~(The NANOGrav
  Collaboration)}}}, \bibinfo{journal}{\apj} \textbf{\bibinfo{volume}{821}},
  \bibinfo{eid}{13} (\bibinfo{year}{2016}), \eprint{1508.03024}.

\bibitem[{\citenamefont{{Milosavljevi{\'c}} and
  {Merritt}}(2003)}]{2003AIPC..686..201M}
\bibinfo{author}{\bibfnamefont{M.}~\bibnamefont{{Milosavljevi{\'c}}}}
  \bibnamefont{and}
  \bibinfo{author}{\bibfnamefont{D.}~\bibnamefont{{Merritt}}}, in
  \emph{\bibinfo{booktitle}{The Astrophysics of Gravitational Wave Sources}},
  edited by \bibinfo{editor}{\bibfnamefont{J.~M.} \bibnamefont{{Centrella}}}
  (\bibinfo{year}{2003}), vol. \bibinfo{volume}{686} of
  \emph{\bibinfo{series}{American Institute of Physics Conference Series}}, pp.
  \bibinfo{pages}{201--210}, \eprint{astro-ph/0212270}.

\bibitem[{\citenamefont{{McWilliams} et~al.}(2012)\citenamefont{{McWilliams},
  {Ostriker}, and {Pretorius}}}]{2012arXiv1211.5377M}
\bibinfo{author}{\bibfnamefont{S.~T.} \bibnamefont{{McWilliams}}},
  \bibinfo{author}{\bibfnamefont{J.~P.} \bibnamefont{{Ostriker}}},
  \bibnamefont{and}
  \bibinfo{author}{\bibfnamefont{F.}~\bibnamefont{{Pretorius}}},
  \bibinfo{journal}{ArXiv e-prints}  (\bibinfo{year}{2012}),
  \eprint{1211.5377}.

\bibitem[{\citenamefont{{Ravi} et~al.}(2014)\citenamefont{{Ravi}, {Wyithe},
  {Shannon}, {Hobbs}, and {Manchester}}}]{2014MNRAS.442...56R}
\bibinfo{author}{\bibfnamefont{V.}~\bibnamefont{{Ravi}}},
  \bibinfo{author}{\bibfnamefont{J.~S.~B.} \bibnamefont{{Wyithe}}},
  \bibinfo{author}{\bibfnamefont{R.~M.} \bibnamefont{{Shannon}}},
  \bibinfo{author}{\bibfnamefont{G.}~\bibnamefont{{Hobbs}}}, \bibnamefont{and}
  \bibinfo{author}{\bibfnamefont{R.~N.} \bibnamefont{{Manchester}}},
  \bibinfo{journal}{\mnras} \textbf{\bibinfo{volume}{442}}, \bibinfo{pages}{56}
  (\bibinfo{year}{2014}), \eprint{1404.5183}.

\bibitem[{\citenamefont{{Vasiliev} et~al.}(2014)\citenamefont{{Vasiliev},
  {Antonini}, and {Merritt}}}]{2014ApJ...785..163V}
\bibinfo{author}{\bibfnamefont{E.}~\bibnamefont{{Vasiliev}}},
  \bibinfo{author}{\bibfnamefont{F.}~\bibnamefont{{Antonini}}},
  \bibnamefont{and}
  \bibinfo{author}{\bibfnamefont{D.}~\bibnamefont{{Merritt}}},
  \bibinfo{journal}{\apj} \textbf{\bibinfo{volume}{785}}, \bibinfo{eid}{163}
  (\bibinfo{year}{2014}), \eprint{1311.1167}.

\bibitem[{\citenamefont{{Sampson} et~al.}(2015)\citenamefont{{Sampson},
  {Cornish}, and {McWilliams}}}]{2015PhRvD..91h4055S}
\bibinfo{author}{\bibfnamefont{L.}~\bibnamefont{{Sampson}}},
  \bibinfo{author}{\bibfnamefont{N.~J.} \bibnamefont{{Cornish}}},
  \bibnamefont{and} \bibinfo{author}{\bibfnamefont{S.~T.}
  \bibnamefont{{McWilliams}}}, \bibinfo{journal}{\prd}
  \textbf{\bibinfo{volume}{91}}, \bibinfo{eid}{084055} (\bibinfo{year}{2015}),
  \eprint{1503.02662}.

\bibitem[{\citenamefont{{Phinney}}(2001)}]{2001astro.ph..8028P}
\bibinfo{author}{\bibfnamefont{E.~S.} \bibnamefont{{Phinney}}},
  \bibinfo{journal}{ArXiv Astrophysics e-prints}  (\bibinfo{year}{2001}),
  \eprint{astro-ph/0108028}.

\bibitem[{\citenamefont{{Peters} and {Mathews}}(1963)}]{1963PhRv..131..435P}
\bibinfo{author}{\bibfnamefont{P.~C.} \bibnamefont{{Peters}}} \bibnamefont{and}
  \bibinfo{author}{\bibfnamefont{J.}~\bibnamefont{{Mathews}}},
  \bibinfo{journal}{Physical Review} \textbf{\bibinfo{volume}{131}},
  \bibinfo{pages}{435} (\bibinfo{year}{1963}).

\bibitem[{\citenamefont{{Anholm} et~al.}(2009)\citenamefont{{Anholm},
  {Ballmer}, {Creighton}, {Price}, and {Siemens}}}]{2009PhRvD..79h4030A}
\bibinfo{author}{\bibfnamefont{M.}~\bibnamefont{{Anholm}}},
  \bibinfo{author}{\bibfnamefont{S.}~\bibnamefont{{Ballmer}}},
  \bibinfo{author}{\bibfnamefont{J.~D.~E.} \bibnamefont{{Creighton}}},
  \bibinfo{author}{\bibfnamefont{L.~R.} \bibnamefont{{Price}}},
  \bibnamefont{and}
  \bibinfo{author}{\bibfnamefont{X.}~\bibnamefont{{Siemens}}},
  \bibinfo{journal}{\prd} \textbf{\bibinfo{volume}{79}}, \bibinfo{eid}{084030}
  (\bibinfo{year}{2009}), \eprint{0809.0701}.

\bibitem[{\citenamefont{{Siemens} et~al.}(2013)\citenamefont{{Siemens},
  {Ellis}, {Jenet}, and {Romano}}}]{2013CQGra..30v4015S}
\bibinfo{author}{\bibfnamefont{X.}~\bibnamefont{{Siemens}}},
  \bibinfo{author}{\bibfnamefont{J.}~\bibnamefont{{Ellis}}},
  \bibinfo{author}{\bibfnamefont{F.}~\bibnamefont{{Jenet}}}, \bibnamefont{and}
  \bibinfo{author}{\bibfnamefont{J.~D.} \bibnamefont{{Romano}}},
  \bibinfo{journal}{Classical and Quantum Gravity}
  \textbf{\bibinfo{volume}{30}}, \bibinfo{eid}{224015} (\bibinfo{year}{2013}),
  \eprint{1305.3196}.

\bibitem[{\citenamefont{{Chamberlin} et~al.}(2015)\citenamefont{{Chamberlin},
  {Creighton}, {Siemens}, {Demorest}, {Ellis}, {Price}, and
  {Romano}}}]{2015PhRvD..91d4048C}
\bibinfo{author}{\bibfnamefont{S.~J.} \bibnamefont{{Chamberlin}}},
  \bibinfo{author}{\bibfnamefont{J.~D.~E.} \bibnamefont{{Creighton}}},
  \bibinfo{author}{\bibfnamefont{X.}~\bibnamefont{{Siemens}}},
  \bibinfo{author}{\bibfnamefont{P.}~\bibnamefont{{Demorest}}},
  \bibinfo{author}{\bibfnamefont{J.}~\bibnamefont{{Ellis}}},
  \bibinfo{author}{\bibfnamefont{L.~R.} \bibnamefont{{Price}}},
  \bibnamefont{and} \bibinfo{author}{\bibfnamefont{J.~D.}
  \bibnamefont{{Romano}}}, \bibinfo{journal}{\prd}
  \textbf{\bibinfo{volume}{91}}, \bibinfo{eid}{044048} (\bibinfo{year}{2015}),
  \eprint{1410.8256}.

\bibitem[{\citenamefont{{Dotti} et~al.}(2007)\citenamefont{{Dotti}, {Colpi},
  {Haardt}, and {Mayer}}}]{2007MNRAS.379..956D}
\bibinfo{author}{\bibfnamefont{M.}~\bibnamefont{{Dotti}}},
  \bibinfo{author}{\bibfnamefont{M.}~\bibnamefont{{Colpi}}},
  \bibinfo{author}{\bibfnamefont{F.}~\bibnamefont{{Haardt}}}, \bibnamefont{and}
  \bibinfo{author}{\bibfnamefont{L.}~\bibnamefont{{Mayer}}},
  \bibinfo{journal}{\mnras} \textbf{\bibinfo{volume}{379}},
  \bibinfo{pages}{956} (\bibinfo{year}{2007}), \eprint{astro-ph/0612505}.

\bibitem[{\citenamefont{{Di Matteo} et~al.}(2001)\citenamefont{{Di Matteo},
  {Carilli}, and {Fabian}}}]{2001ApJ...547..731D}
\bibinfo{author}{\bibfnamefont{T.}~\bibnamefont{{Di Matteo}}},
  \bibinfo{author}{\bibfnamefont{C.~L.} \bibnamefont{{Carilli}}},
  \bibnamefont{and} \bibinfo{author}{\bibfnamefont{A.~C.}
  \bibnamefont{{Fabian}}}, \bibinfo{journal}{\apj}
  \textbf{\bibinfo{volume}{547}}, \bibinfo{pages}{731} (\bibinfo{year}{2001}),
  \eprint{astro-ph/0005516}.

\bibitem[{\citenamefont{{Armitage} and
  {Natarajan}}(2002)}]{2002ApJ...567L...9A}
\bibinfo{author}{\bibfnamefont{P.~J.} \bibnamefont{{Armitage}}}
  \bibnamefont{and}
  \bibinfo{author}{\bibfnamefont{P.}~\bibnamefont{{Natarajan}}},
  \bibinfo{journal}{\apjl} \textbf{\bibinfo{volume}{567}}, \bibinfo{pages}{L9}
  (\bibinfo{year}{2002}), \eprint{astro-ph/0201318}.

\bibitem[{\citenamefont{{Dotti} et~al.}(2015)\citenamefont{{Dotti}, {Merloni},
  and {Montuori}}}]{2015MNRAS.448.3603D}
\bibinfo{author}{\bibfnamefont{M.}~\bibnamefont{{Dotti}}},
  \bibinfo{author}{\bibfnamefont{A.}~\bibnamefont{{Merloni}}},
  \bibnamefont{and}
  \bibinfo{author}{\bibfnamefont{C.}~\bibnamefont{{Montuori}}},
  \bibinfo{journal}{\mnras} \textbf{\bibinfo{volume}{448}},
  \bibinfo{pages}{3603} (\bibinfo{year}{2015}), \eprint{1502.03101}.

\bibitem[{\citenamefont{{Goicovic} et~al.}(2016)\citenamefont{{Goicovic},
  {Cuadra}, {Sesana}, {Stasyszyn}, {Amaro-Seoane}, and
  {Tanaka}}}]{2016MNRAS.455.1989G}
\bibinfo{author}{\bibfnamefont{F.~G.} \bibnamefont{{Goicovic}}},
  \bibinfo{author}{\bibfnamefont{J.}~\bibnamefont{{Cuadra}}},
  \bibinfo{author}{\bibfnamefont{A.}~\bibnamefont{{Sesana}}},
  \bibinfo{author}{\bibfnamefont{F.}~\bibnamefont{{Stasyszyn}}},
  \bibinfo{author}{\bibfnamefont{P.}~\bibnamefont{{Amaro-Seoane}}},
  \bibnamefont{and} \bibinfo{author}{\bibfnamefont{T.~L.}
  \bibnamefont{{Tanaka}}}, \bibinfo{journal}{\mnras}
  \textbf{\bibinfo{volume}{455}}, \bibinfo{pages}{1989} (\bibinfo{year}{2016}),
  \eprint{1507.05596}.

\bibitem[{\citenamefont{{Z.~Arzoumanian et al.~(The NANOGrav
  Collaboration)}}(2015)}]{2015ApJ...813...65T}
\bibinfo{author}{\bibnamefont{{Z.~Arzoumanian et al.~(The NANOGrav
  Collaboration)}}}, \bibinfo{journal}{\apj} \textbf{\bibinfo{volume}{813}},
  \bibinfo{eid}{65} (\bibinfo{year}{2015}), \eprint{1505.07540}.

\bibitem[{\citenamefont{Olver et~al.}(2010)\citenamefont{Olver, Lozier,
  Boisvert, and Clark}}]{NHMF}
\bibinfo{editor}{\bibfnamefont{F.~W.~J.} \bibnamefont{Olver}},
  \bibinfo{editor}{\bibfnamefont{D.~W.} \bibnamefont{Lozier}},
  \bibinfo{editor}{\bibfnamefont{R.~F.} \bibnamefont{Boisvert}},
  \bibnamefont{and} \bibinfo{editor}{\bibfnamefont{C.~W.} \bibnamefont{Clark}},
  eds., \emph{\bibinfo{title}{{NIST Handbook of Mathematical Functions}}}
  (\bibinfo{publisher}{Cambridge University Press}, \bibinfo{address}{New York,
  NY}, \bibinfo{year}{2010}).

\bibitem[{\citenamefont{{Sesana}}(2013)}]{2013MNRAS.433L...1S}
\bibinfo{author}{\bibfnamefont{A.}~\bibnamefont{{Sesana}}},
  \bibinfo{journal}{\mnras} \textbf{\bibinfo{volume}{433}}, \bibinfo{pages}{L1}
  (\bibinfo{year}{2013}), \eprint{1211.5375}.

\bibitem[{\citenamefont{{Ravi} et~al.}(2015)\citenamefont{{Ravi}, {Wyithe},
  {Shannon}, and {Hobbs}}}]{2015MNRAS.447.2772R}
\bibinfo{author}{\bibfnamefont{V.}~\bibnamefont{{Ravi}}},
  \bibinfo{author}{\bibfnamefont{J.~S.~B.} \bibnamefont{{Wyithe}}},
  \bibinfo{author}{\bibfnamefont{R.~M.} \bibnamefont{{Shannon}}},
  \bibnamefont{and} \bibinfo{author}{\bibfnamefont{G.}~\bibnamefont{{Hobbs}}},
  \bibinfo{journal}{\mnras} \textbf{\bibinfo{volume}{447}},
  \bibinfo{pages}{2772} (\bibinfo{year}{2015}), \eprint{1406.5297}.

\bibitem[{\citenamefont{{Rasskazov} and {Merritt}}(2016)}]{2016arXiv160607484R}
\bibinfo{author}{\bibfnamefont{A.}~\bibnamefont{{Rasskazov}}} \bibnamefont{and}
  \bibinfo{author}{\bibfnamefont{D.}~\bibnamefont{{Merritt}}},
  \bibinfo{journal}{ArXiv e-prints}  (\bibinfo{year}{2016}),
  \eprint{1606.07484}.

\bibitem[{\citenamefont{{Sesana} et~al.}(2016)\citenamefont{{Sesana},
  {Shankar}, {Bernardi}, and {Sheth}}}]{2016MNRAS.463L...6S}
\bibinfo{author}{\bibfnamefont{A.}~\bibnamefont{{Sesana}}},
  \bibinfo{author}{\bibfnamefont{F.}~\bibnamefont{{Shankar}}},
  \bibinfo{author}{\bibfnamefont{M.}~\bibnamefont{{Bernardi}}},
  \bibnamefont{and} \bibinfo{author}{\bibfnamefont{R.~K.}
  \bibnamefont{{Sheth}}}, \bibinfo{journal}{\mnras}
  \textbf{\bibinfo{volume}{463}}, \bibinfo{pages}{L6} (\bibinfo{year}{2016}),
  \eprint{1603.09348}.

\bibitem[{\citenamefont{{Kelley} et~al.}(2016)\citenamefont{{Kelley}, {Blecha},
  and {Hernquist}}}]{2016arXiv160601900K}
\bibinfo{author}{\bibfnamefont{L.~Z.} \bibnamefont{{Kelley}}},
  \bibinfo{author}{\bibfnamefont{L.}~\bibnamefont{{Blecha}}}, \bibnamefont{and}
  \bibinfo{author}{\bibfnamefont{L.}~\bibnamefont{{Hernquist}}},
  \bibinfo{journal}{ArXiv e-prints}  (\bibinfo{year}{2016}),
  \eprint{1606.01900}.

\end{thebibliography}

\end{document}